\documentclass[aps,showpacs,pre,superscriptaddress,nofootinbib,notitlepage,a4paper]{revtex4}
\usepackage[english]{babel}
\usepackage{dsfont}
\usepackage{mathtools,nccmath}
\usepackage{amsmath}
\usepackage{amsfonts}
\usepackage{amssymb}
\usepackage{xcolor}
\usepackage{graphicx}
\usepackage{psfrag}
\usepackage{float}
\usepackage{placeins}
\usepackage{scalerel}

\usepackage{tikz}
\usetikzlibrary{patterns,decorations,arrows,shapes.geometric,arrows,mindmap,backgrounds,positioning,positioning,fit,backgrounds,positioning,arrows,matrix,calc}
\usetikzlibrary{shapes,backgrounds,mindmap,shadows,shapes.geometric,topaths,snakes,arrows,pgfplots.groupplots,fadings,calc,decorations.markings,positioning,chains,fit}
\usetikzlibrary{arrows,shapes,positioning,quotes}
\usetikzlibrary{decorations.markings}
\tikzstyle arrowstyle=[scale=1]
\tikzstyle directed=[postaction={decorate,decoration={markings,
    mark=at position .65 with {\arrow[arrowstyle]{stealth}}}}]
\usetikzlibrary{shapes.geometric}
\tikzset{mynode/.style={draw, shape=circle, fill=black, color=black, line width=0pt, inner sep=2.5pt, label position=0, text=black}}
\tikzset{myedge/.style={draw,thick,solid}}
\tikzset{bicoloredge/.style={dashed, dash pattern=on 4pt off 4pt, blue, postaction={draw, dashed, dash pattern=on 4pt off 4pt, red, dash phase=4pt}, ultra thick}}
\usepackage{circuitikz}
\usepackage{ifthen}
\usepackage[hidelinks]{hyperref}
\usepackage{adjustbox}

\newcommand{\beq}{\begin{equation}} 
\newcommand{\eeq}{\end{equation}}
\newcommand{\bea}{\begin{align}}
\newcommand{\eea}{\end{align}}

\newcommand{\hpi}{\widehat{\pi}}
\newcommand{\hf}{\widehat{f}}
\newcommand{\hL}{\widehat{L}}
\newcommand{\halpha}{\widehat{\alpha}}
\newcommand{\hchi}{\widehat{\chi}}
\newcommand{\hpsi}{\widehat{\psi}}
\newcommand{\hA}{\widehat{A}}
\newcommand{\tL}{\widetilde{L}}
\newcommand{\tchi}{\widetilde{\chi}}
\newcommand{\tf}{\widetilde{f}}

\newcommand{\E}{\mathbb{E}}
\DeclareMathOperator{\ov}{\text{ov}}
\newcommand{\one}{\mathds{1}}

\newcommand{\pzero}[2]{\ifthenelse{\equal{#2}{}}{\mathbb{P}_0^{(#1)}}{\mathbb{P}_0^{(#1)}[#2]}}
\newcommand{\pone}[2]{\ifthenelse{\equal{#2}{}}{\mathbb{P}_1^{(#1)}}{\mathbb{P}_1^{(#1)}[#2]}}
\DeclareMathOperator{\Po}{\text{Po}}

\begin{document}

\bibliographystyle{myunsrt}

\title{Faster algorithms for the alignment of sparse correlated Erd\H os-R\'enyi random graphs}

\author{Andrea Muratori}
\affiliation{Dipartimento di Fisica e Astronomia, Universit\`a di Bologna, 40127 Bologna, Italy}

\author{Guilhem Semerjian}
\affiliation{Laboratoire de physique de l'\'Ecole Normale Sup\'erieure, ENS, Universit\'e PSL, CNRS, Sorbonne Universit\'e, Universit\'e Paris Cit\'e, F-75005 Paris, France} 

\begin{abstract}
The correlated Erd\H os-R\'enyi random graph ensemble is a probability law on pairs of graphs with $n$ vertices, parametrized by their average degree $\lambda$ and their correlation coefficient $s$. It can be used as a benchmark for the graph alignment problem, in which the labels of the vertices of one of the graphs are reshuffled by an unknown permutation; the goal is to infer this permutation and thus properly match the pairs of vertices in both graphs. A series of recent works has unveiled the role of Otter's constant $\alpha$ (that controls the exponential rate of growth of the number of unlabeled rooted trees as a function of their sizes) in this problem: for $s>\sqrt{\alpha}$ and $\lambda$ large enough it is possible to recover in a time polynomial in $n$ a positive fraction of the hidden permutation. The exponent of this polynomial growth is however quite large and depends on the other parameters, which limits the range of applications of the algorithm. In this work we present a family of faster algorithms for this task, show through numerical simulations that their accuracy is only slightly reduced with respect to the original one, and conjecture that they undergo, in the large $\lambda$ limit, phase transitions at modified Otter's thresholds $\sqrt{\widehat{\alpha}}>\sqrt{\alpha}$, with $\widehat{\alpha}$ related to the enumeration of a restricted family of trees.
\end{abstract}

\maketitle

\tableofcontents

\section{Introduction}

In graph alignment problems one is given a pair of graphs and has to find a correspondence between the vertices of those two, based on the observation of their structures. These problems have applications in several fields of data analysis, in particular de-anonymization of networks~\cite{narayanan_-anonymizing_2009,agrawal_differential_2008,pedarsani_privacy_2011}, alignment of protein interaction networks or molecular networks~\cite{berg_cross-species_2006,li_alignment_2007,ramani_exploiting_2003,singh_global_2008}, pattern recognition and image processing~\cite{berg_shape_2005,conte_thirty_2004}. They are also used as benchmarks for graph neural networks architectures~\cite{azizian_expressive_2021,nowak_revised_2018}.

In absence of prior information on the graphs, one possible objective is to maximize the number of common edges in the aligned graphs; more formally, given a pair of graphs on $n$ vertices, represented by their adjacency matrices $A,B\in\mathbb{R}^{n\times n}$, the goal is to find the permutation $\pi\in \mathcal{S}_n$ ($\mathcal{S}_n$ being the symmetric group of $n$ elements) between their vertices that maximizes $\sum_{i<j} A_{i,j}B_{\pi(i),\pi(j)}$. This optimization task is called a quadratic assignment problem~\cite{du_quadratic_1998}, which is NP-hard in general.

Our work is situated instead in a Bayesian line of research, in which one postulates a prior distribution on the observed pair of graphs involving a ground truth hidden permutation $\pi_\star$, the goal being to extract as much as possible of information about $\pi_\star$ from the observation of the graphs. Following several previous works we will focus our attention on the correlated Erd\H{o}s-R\'enyi (ER) random graph ensemble with parameters $(\lambda,s)$. In a pair of graphs $(G,H)$ drawn from this ensemble, $G$ and $H$ have individually the law of a usual ER random graph, i.e. each of their edges is present with probability $\lambda/n$, but they are correlated, the parameter $s \in [0,1]$ controlling the probability of a common edge in both graphs. One draws in addition a uniformly random permutation $\pi_\star$, creates a graph $G'$ by relabeling the vertices of $H$ through $\pi_\star$, and provides an observer with the pair $(G,G')$. Various statistical tasks can then be defined, for instance the hypothesis testing problem of distinguishing whether $(G,H)$ are independent or correlated, or the recovery problem of inferring $\pi_\star$ from the observation of $(G,G')$. The latter recovery task can be subdivided according to the level of accuracy required on the reconstruction of $\pi_\star$: in its exact recovery version all vertices have to be correctly matched between the two graphs, the almost exact version allows for a vanishing fraction of mistakes (asymptotically in the large $n$ limit), while weak (or partial) recovery only demands a positive fraction of $\pi_\star$ to be correctly inferred. One can furthermore investigate the existence of polynomial-time algorithms to perform these various tasks, so-called statistical-computational gaps appearing when a problem is known to be information-theoretically feasible but no efficient algorithm has been discovered yet.

\begin{figure}
\begin{tikzpicture}
\draw (1.4,.8) node {impossible};
\draw (4.2,2.4) node {easy};
\draw (5.5,.8) node {hard};
\draw[dashed] (5,1.55)--(6.4,1.55);
\draw  (6.4,1.5) node [right] {$\sqrt{\alpha}$};
\draw[dashed] (5,1.85)--(6.4,1.85);
\draw  (6.4,1.95) node [right] {$\sqrt{\halpha}$};
\draw[-latex] (0,0) -- (0,3.5) node [left] {$s$};
\draw[-latex] (0,0) -- (6.5,0) node [below] {$\lambda$};
\draw (6.3,3) -- (0,3) node [left] {$1$};
\draw (1.5,3) to[out=-65,in=175] (6.3,.2);
\draw (1.8,2.45) to[out=-30,in=180] (6.,1.6);
\draw (1.5,3) node [above] {$1$};
\fill[black] (1.8,2.45) circle (1.2pt);
\draw[dotted] (1.5,3) to[out=-45,in=180] (6.,1.9);
\end{tikzpicture}
\caption{A sketch of the phase diagram for the partial recovery task in the correlated sparse Erd\H os-R\'enyi ensemble with finite average degree $\lambda$ and correlation $s$. The information-theoretic impossible region corresponds to $\lambda s <1$; its complement is divided in an easy phase with boundary $s_{\rm algo}(\lambda)$ where polynomial-time algorithms are known to exist, and a hard phase where they are yet to be discovered (or proven not to exist). The dotted line represents the conjectured algorithmic phase transition line $\widehat{s}_{\rm algo}(\lambda)$ for the procedures introduced in the present work.}
\label{fig_sketch_pd}
\end{figure}
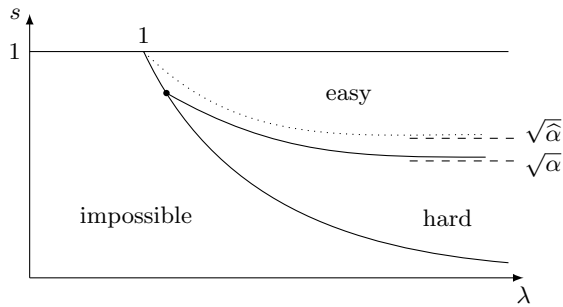

These questions have been thoroughly explored for the correlated ER ensemble in many previous works, unveiling a rich phenomenology depending on the parameters $(\lambda,s)$ of the problem. The information-theoretic limits for the exact and almost exact recovery have been established in~\cite{cullina_improved_2016,wu_settling_2022} and shown to lie in the dense regime where $\lambda$ diverges with $n$; efficient algorithms for this task have been put forward in~\cite{ding_efficient_2018,fan_spectral_2019,fan_spectral_2019_2,mao_exact_2022,mao_random_2023,DiLi23}, for large enough correlations $s$. We will concentrate in this work on the sparse regime with $\lambda$ fixed in the large $n$ limit (see Fig.~\ref{fig_sketch_pd} for a sketch of the phase diagram that summarizes the following statements), in which case the relevant task is the partial recovery one. Its information-theoretic limit is now known to be $\lambda s=1$: the impossibility of recovering more than a vanishing fraction of $\pi_\star$ when $\lambda s < 1$ has been established in~\cite{pmlr-v134-ganassali21a}, while it was shown in~\cite{ding_matching_2022} that $\lambda s > 1$ is a sufficient condition for the possibility of weak recovery (improving previous results from~\cite{hall_partial_2022,wu_settling_2022}). Polynomial-time algorithms that achieves partial recovery are at present only known in a portion of the $\lambda s > 1$ region, labeled ``easy'' on the sketch of Fig.~\ref{fig_sketch_pd}~\cite{ganassali_tree_2020,ganassali_correlation_2022,piccioli_aligning_2022,mao_random_2023}; the status of the remaining ``hard'' phase is less clear than in other inference problems~\cite{zdeborova_statistical_2016}, since for these parameters we have neither efficient procedures nor hardness results ruling out their existences under some computational complexity hypotheses or within some family of algorithms~\cite{GaMoZd22,bandeira2022franz} (see however~\cite{DiDuLi23} for hardness results on the related detection problem). 

Our work will build on a message-passing algorithm introduced in~\cite{ganassali_correlation_2022,piccioli_aligning_2022}, that computes a similarity score between the vertices of $G$ and $G'$ based on the exploration of their local neighborhoods. It succeeds in the partial recovery task for $s$ larger than some threshold $s_{\rm algo}(\lambda)$, the lower boundary of the easy phase in Fig.~\ref{fig_sketch_pd}. This function $s_{\rm algo}(\lambda)$ is not known analytically; some bounds on it have been established in~\cite{ganassali_correlation_2022}, it was estimated through numerical simulations in~\cite{piccioli_aligning_2022}, and its large $\lambda$ limit was proven in~\cite{ganassali_statistical_2022} to be equal to $\sqrt{\alpha} \approx 0.581$, where $\alpha$ is the Otter's constant~\cite{otter_number_1948} (defined in terms of the enumeration of a family of trees). This Otter's threshold first appeared in the context of graph alignment problems in~\cite{mao_testing_2022}, for the correlation detection task, and is also the large degree limit of the algorithmic threshold of a different procedure for partial recovery presented in~\cite{mao_random_2023} (that also applies in the dense regime to perform exact recovery). 

The message-passing algorithm of~\cite{ganassali_correlation_2022,piccioli_aligning_2022} runs in a time polynomial in the size $n$ of the graphs, but with an exponent that depends on $\lambda$, which limits in practice its potential use to rather small values of $\lambda$ (or of $n$). Our contribution in the present work will be the introduction of a family of simplified similarity scores, depending on an integer parameter $m$ and inspired by the analytical results of~\cite{ganassali_statistical_2022}, which can be computed in a much faster way while preserving most of the accuracy of the original algorithm. The parameter $m$ controls the number of terms kept in the truncation of the similarity score formula of~\cite{ganassali_correlation_2022,piccioli_aligning_2022}, in such a way that the running time of our algorithm scales asymptotically with $n$ and $\lambda$ as $O(n^2 (\log n)\lambda^{c(m)})$. For the first two non-trivial values of $m$, $m=2$ and $m=3$ on which we present more detailed results, the exponent $c(m)=2$. According to the numerical simulations we shall present, the ($m$-dependent) algorithmic threshold $\widehat{s}_{\rm algo}(\lambda)$ above which these simplified procedures succeed in the partial recovery task (plotted as a dotted line in Fig.~\ref{fig_sketch_pd}) is only slightly larger than $s_{\rm algo}(\lambda)$, which suggests a satisfying compromise between the loss in statistical accuracy and the gain in computational speed. Furthermore, we conjecture, based on some analytical computations, that the large $\lambda$ limit of $\widehat{s}_{\rm algo}(\lambda)$ is given by a modified Otter's threshold $\sqrt{\halpha}$, with $\halpha$ related to the enumeration of a restricted set of trees. For the two smallest values of $m$, that correspond to the fastest computations of the simplified scores, these thresholds are~\cite{otter_number_1948} $\sqrt{\halpha} \approx 0.635$ and $\sqrt{\halpha} \approx 0.596$, the last one being rather close to the original threshold $\sqrt{\alpha} \approx 0.581$.

The rest of the paper is organized as follows. In Section~\ref{sec_MPA} we define precisely the graph alignment problem, review the message-passing algorithm previously studied in~\cite{ganassali_correlation_2022,piccioli_aligning_2022}, and introduce its simplified variants. The performances of the latter are characterized through numerical simulations in Sec.~\ref{sec:numerical-results}, while their expected behavior in the thermodynamic limit is discussed in Sec.~\ref{sec:trees-results}. Some conclusions are drawn in Section~\ref{sec_conclu}, more technical details being deferred to several Appendices (in particular App.~\ref{app:proof-truncation} and \ref{app:martingale} contain the proofs of additional properties of the orthogonal polynomials on trees introduced in~\cite{ganassali_statistical_2022}, that might be useful in a broader context). 

\section{Message-passing algorithms for the graph alignment problem}
\label{sec_MPA}

\subsection{Basic definitions}
\label{sec_basic_def}

For a positive integer $n$ we shall denote $[n]=\{1,\dots,n\}$ the set of the first $n$ integers. 

A graph $G=(V,E)$ is given by a set $V$ of vertices and a set $E$ of edges, where each edge is an unordered pair $\{i,j\}$ of distinct vertices. Two vertices $i$ and $j$ are said to be adjacent, or neighbors, if $\{i,j\} \in E$. For a vertex $i \in V$ we shall denote $\partial i$ the set of the vertices adjacent to it, and call $d_i=|\partial i|$ its degree, i.e. the number of edges in which it appears as an endvertex. A graph is said to be connected if for any pair of vertices $i,j$ one can find a path of adjacent vertices connecting them; the distance between $i$ and $j$ is then defined as the minimal length of such a path, where the length is the number of edges crossed along the path.

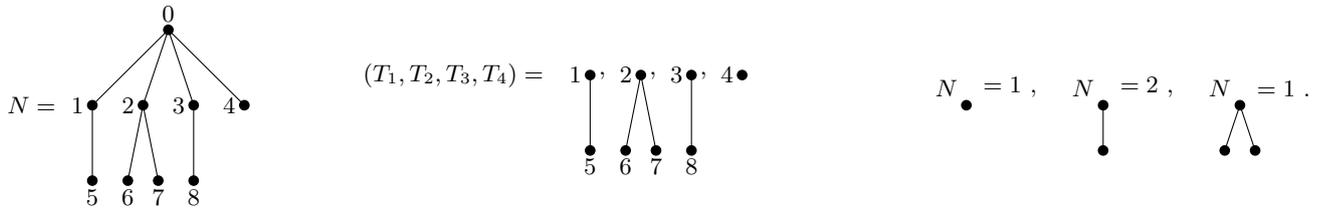
\begin{figure}
\centering
\begin{tikzpicture}
\fill[black] (0,0) circle (2pt) node [above] {$0$};
\fill[black] (-1,-1) circle (2pt) node [left] {$1$};
\fill[black] (-1,-2) circle (2pt) node [below] {$5$};
\fill[black] (-.333,-1) circle (2pt) node [left] {$2$};
\fill[black] (-.533,-2) circle (2pt) node [below] {$6$};
\fill[black] (-.133,-2) circle (2pt) node [below] {$7$};
\fill[black] (.333,-1) circle (2pt) node [left] {$3$};
\fill[black] (.333,-2) circle (2pt) node [below] {$8$};
\fill[black] (1,-1) circle (2pt) node [left] {$4$};
\draw (-1,-2) -- (-1,-1) -- (0,0) -- (1,-1) ;
\draw (0,0) -- (-.333,-1);
\draw (-.533,-2) -- (-.333,-1) -- (-.133,-2);
\draw (.333,-2) -- (.333,-1) -- (0,0);
\draw (-1.8,-1) node {$N=$};
\begin{scope}[xshift=6.55cm,yshift=0.4cm]
\draw (-2.8,-1) node {$(T_1,T_2,T_3,T_4) =$};
\fill[black] (-1,-1) circle (2pt) node [left] {$1$};
\draw (-1,-1) node [right] {$,$};
\fill[black] (-1,-2) circle (2pt) node [below] {$5$};
\fill[black] (-.333,-1) circle (2pt) node [left] {$2$};
\draw (-.333,-1) node [right] {$,$};
\fill[black] (-.533,-2) circle (2pt) node [below] {$6$};
\fill[black] (-.133,-2) circle (2pt) node [below] {$7$};
\fill[black] (.333,-1) circle (2pt) node [left] {$3$};
\draw (.333,-1) node [right] {$,$};
\fill[black] (.333,-2) circle (2pt) node [below] {$8$};
\fill[black] (1,-1) circle (2pt) node [left] {$4$};
\draw (-1,-2) -- (-1,-1);
\draw (-.533,-2) -- (-.333,-1) -- (-.133,-2);
\draw (.333,-2) -- (.333,-1);
\end{scope}
\begin{scope}[xshift=10.5cm,yshift=-1cm]
\fill[black] (0,0) circle (2pt);
\draw (0,0) node [above left] {$N$};
\draw (0,0) node [above right] {$\ =1 \ , $};
\fill[black] (1.8,0) circle (2pt);
\fill[black] (1.8,-.6) circle (2pt);
\draw (1.8,-.6) -- (1.8,0);
\draw (1.8,0) node [above left] {$N$};
\draw (1.8,0) node [above right] {$\ =2 \ , $};

\fill[black] (3.6,0) circle (2pt);
\fill[black] (3.4,-.6) circle (2pt);
\fill[black] (3.8,-.6) circle (2pt);
\draw (3.4,-.6) -- (3.6,0) -- (3.8,-.6);
\draw (3.6,0) node [above left] {$N$};
\draw (3.6,0) node [above right] {$\ =1 \ . $};
\end{scope}
\end{tikzpicture}
\caption{An example of a rooted tree $N$ whose root has $l=4$ offsprings, its decomposition as a list of subtrees, and its unlabeled version defined in terms of the number of copies of unlabeled subtrees rooted at the offsprings of the root.}
\label{fig_def_trees}
\end{figure}

A tree is a non-empty connected graph that is acyclic, in the sense that it does not contain any closed loop of distinct vertices. A rooted tree is a tree in which one vertex is distinguished and called the root; in the graphical representations we shall place the root at the top of the tree. The depth of a vertex $i$ in a rooted tree is the distance between $i$ and the root; the depth of a rooted tree is the maximal depth of its vertices. In a rooted tree, a vertex of depth $d>0$ (hence distinct from the root) has one neighbor of depth $d-1$, called its ancestor, and possibly neighbors of depth $d+1$, called its offsprings, or descendants; its degree is thus its number of offsprings plus one. The size of a tree $T$, denoted $|T|$, is defined as the number of its vertices. Because of their acyclic character, rooted trees admit naturally a recursive decomposition: one can describe a rooted tree $N$ in terms of the degree $l$ of its root, complemented by a list $T_1,\dots,T_l$ of the subtrees rooted at the offsprings of the root (see Fig.~\ref{fig_def_trees} for an illustration).

It will be sometimes convenient to consider two rooted trees as equivalent if they only differ by a relabeling of their vertices that preserves the parenthood relationships, and to call unlabeled rooted trees the associated equivalence classes. We shall denote $\chi_d$ the set of finite unlabeled rooted trees of depth at most $d$; in this unlabeled context one can describe $N \in \chi_{d+1}$ as a vector $\{N_T\}_{T \in \chi_d}$ of non-negative integers, where $N_T$ indicates the number of copies of the subtree $T$ rooted at the offsprings of the root of $N$ (see Fig.~\ref{fig_def_trees} for an example). Since we consider finite trees, the number of non-zero elements in $\{N_T\}$ is finite. The Otter's constant $\alpha$ is defined from the combinatorial properties of these objects~\cite{otter_number_1948}: calling $A_n$ the number of distinct unlabeled rooted trees of size $n$ (irrespectively of their depth), one has $\frac{1}{n} \ln A_n \to \alpha^{-1}$ as $n \to \infty$, in other words $\alpha$ is the inverse of the exponential growth rate of the number of trees with respect to their sizes.

\subsection{Definition of the inference problem}

Let us denote $G(n,\lambda/n,s)$ the correlated Erd\H os-R\'enyi (ER) random graph ensemble with parameters $\lambda \ge 0$ and $s\in[0,1]$. Its elements are pairs $(G,H)$ of graphs on the same vertex set $V=[n]$, with edge sets denoted $E_G$ and $E_H$ respectively. The latter are generated as follows: independently for each of the $n(n-1)/2$ pairs of vertices $i<j$ the edge $\{i,j\}$ between them is included
\begin{itemize}
\item in $E_G$ and $E_H$ with probability $\lambda s/n$;
\item in $E_G$ but not in $E_H$ with probability $\lambda(1-s)/n$;
\item in $E_H$ but not in $E_G$ with probability $\lambda(1-s)/n$;
\item in none of the graphs with probability $1-\lambda(2-s)/n$.
\end{itemize}
Considered individually, $G$ and $H$ have the law of a usual ER random graph, with each edge present with probability $\lambda/n$, that is the parameter $\lambda$ fixes the edge density or equivalently the average degree of such graphs. The correlation between them is controlled by $s$: when $s=1$ $G$ and $H$ are identical, while when $s=\lambda/n$ they are independent. In the large size (thermodynamic) limit $n\to\infty$ there is a positive correlation between the presence of an edge in the two graphs whenever $s>0$.

We further introduce a graph $G'$ which is a reshuffled version of $H$: a permutation $\pi_\star\in\mathcal{S}_n$ is chosen uniformly at random from the symmetric group $\mathcal{S}_n$ (i.e. the set of permutations of $[n]$), and the edge set $E_{G'}$ is defined as the one of $H$ once vertices are relabeled through $\pi_\star$. More explicitly, $\{i,j\} \in E_H$ if and only if $\{\pi_\star (i), \pi_\star(j)\} \in E_{G'}$.

The graph alignment problem we are interested in consists in inferring the hidden permutation $\pi_\star$ from the sole observation of the graphs $(G,G')$. We place ourselves in a Bayesian perspective, assuming that the observer knows that $(G,G')$ has been generated in the way defined above, as well as the value of the parameters $\lambda$ and $s$. In this setting all the information on the hidden permutation is contained in the posterior distribution,
\begin{equation}
\mathbb{P}(\pi_\star=\pi|G,G')=\frac{\mathbb{P}(\pi_\star=\pi,G,G')}{\mathbb{P}(G,G')}\propto\mathbb{P}(\pi_\star=\pi,G,G') \ ,
\label{eq_posterior}
\end{equation}
where the symbol $\propto$ hides a normalization constant independent on $\pi$. The objective of the inference problem is to construct an estimator $\hpi = \hpi(G,G')$ that approximates $\pi_\star$ as closely as possible. This notion of closeness, and the related Bayes-optimality of the estimator, depends on the constraints imposed on $\hpi$, and on the choice of the measure of the distance between $\hpi$ and $\pi_\star$. Since we are interested in the regime of finite average degree $\lambda$, where one can only hope for a partial recovery of $\pi_\star$, a relevant choice is to consider $\hpi$ to be a function from $[n]$ to $[n]$ (not necessarily a permutation) and to measure the accuracy of the estimation in terms of the overlap
\begin{equation}
\ov(\hpi,\pi_\star)=\frac{1}{n}\sum_{i=1}^{n}\mathbb{I}(\hpi(i)=\pi_\star(i)) \ ,
\label{eq:overlap}
\end{equation}
which counts the fraction of the vertices $i$ of $G$ which are correctly matched to the hidden corresponding vertex $\pi_\star(i)$ of $G'$ (see~\cite{piccioli_aligning_2022} for a discussion of alternative measures of the distance between $\hpi$ and $\pi_\star$). The partial recovery task is thus successfully performed if the overlap remains strictly positive in the large size limit. The optimal estimator which maximizes this overlap, on average with respect to the law of the generation of $(G,G',\pi_\star)$, is then found to be
\begin{equation}
\hpi(G,G')(i) = \arg\max_{i'}\mathbb{P}(\pi_\star(i) = i'|G,G') \ .
\label{eq:argmaxi}
\end{equation}
If one can write rather easily an explicit form of the posterior distribution from equation (\ref{eq_posterior}), the exact computation of the marginal probability in (\ref{eq:argmaxi}) is in general a NP-hard problem that cannot be solved in polynomial time; this motivates the introduction of a simplified algorithmic strategy, detailed in the next subsection.

\subsection{The original algorithm}
\label{sec:original-alg}

We shall now describe the algorithm introduced in~\cite{ganassali_correlation_2022,piccioli_aligning_2022}, that will be the starting point for the present work. Its main idea is to build an estimator $\hpi$ from a simplified version of Eq.~(\ref{eq:argmaxi}), in which part of the information in the posterior distribution is discarded. More precisely, one introduces a depth, or distance, parameter $d$, and denotes $G_{i,d}$ (resp. $G'_{i',d}$) the neighborhood of the vertex $i$ (resp. $i'$) up to distance $d$, that is the subgraph of $G$ (resp. $G'$) induced by the vertices at distance at most $d$ from $i$ (resp. $i'$). Consider now the posterior probability that appeared in Eq.~(\ref{eq:argmaxi}), and simplify it by using only the local information up to distance $d$ around the vertices $i$ and $i'$:
\begin{equation}
\begin{split}
\mathbb{P}(\pi_\star(i) = i'|G_{i,d},G'_{i',d})&=\frac{\mathbb{P}(\pi_\star(i) = i',G_{i,d},G'_{i',d})}{\mathbb{P}(G_{i,d},G'_{i',d})}=\frac{\mathbb{P}(G_{i,d},G'_{i',d}|\pi_\star(i) = i')\mathbb{P}(\pi_\star(i) = i')}{\mathbb{P}(G_{i,d},G'_{i',d})}\\
&=\frac{1}{n}\frac{\mathbb{P}(G_{i,d},G'_{i',d}|\pi_\star(i) = i')}{\mathbb{P}(G_{i,d},G'_{i',d})} = \frac{1}{n} L^{(d)}_{i,i'} \ ,
\label{eq:trunc-post}
\end{split}
\end{equation}
where we used Bayes rule to reverse the conditioning, and the uniformity in the choice of the permutation $\pi_\star$, that implies that the prior probability of the event $\pi_\star(i) = i'$ is $1/n$; the last equality has to be understood as a definition of $L^{(d)}_{i,i'}$. This last quantity can be computed exactly when the thermodynamic limit $n\to\infty$ is taken with a finite value of $d$ (or when $d$ grows slowly with $n$, i.e. $d = \mu \log n$ with $\mu$ a sufficiently small constant). Indeed, in this regime the local neighborhood of a given vertex in a random graph is, with high probability, a random tree. This is a well-known fact in the usual ER ensemble, whose local limit of depth $d$ is the Galton-Watson tree with Poissonian offspring probability, denoted $\pzero{d}{T}$. This distribution is defined as follows: $T$ is a rooted tree whose root has a number $l$ of offsprings drawn from the Poisson law with mean $\lambda$, $\Po(\lambda;l)=e^{-\lambda}\frac{\lambda^l}{l!}$. Each of these offsprings is itself the root of a tree drawn from the law $\pzero{d-1}{}$, this recursive definition being complemented by the boundary condition for $d=0$: $\pzero{0}{}$ is supported on the trivial tree containing only the root vertex. The denominator in the definition of $L^{(d)}_{i,i'}$ thus converges in the thermodynamic limit to $\pzero{d}{G_{i,d}} \pzero{d}{G'_{i',d}}$: in this probability the permutation $\pi_\star$ is chosen uniformly at random, the vertex $i'$ of $G'$ corresponds to a uniformly random vertex of $H$, hence the laws of $G_{i,d}$ and $G'_{i',d}$ are asymptotically independent. On the contrary in the numerator the conditioning on the event $\pi_\star(i) = i'$ induces a strong correlation on the joint law of the neighborhoods $(G_{i,d},G'_{i',d})$, whose roots correspond to the same vertex in the correlated pair of graphs $(G,H)$ (before the reshuffling of the vertex labels by $\pi_\star$). We shall denote $\pone{d}{T,T'}$ the limit of this joint law in the thermodynamic limit; generalizing the Galton-Watson construction to this correlated case one can obtain a recursive description of $\pone{d}{}$, by induction on the depth $d$.  It will be convenient in the following to write down the induction on the so-called likelihood ratio
\beq
L^{(d)}(T,T')=\frac{\mathbb{P}_1^{(d)}[T,T']}{\mathbb{P}_0^{(d)}[T]\mathbb{P}_0^{(d)}[T']} \ .
\label{eq_def_L}
\eeq
Let us consider a pair $(N,N')$ of trees of depth at most $d+1$; we shall denote $l$ (resp. $l'$) the number of offsprings of the root of $N$ (resp. $N'$), and $T_1,\dots,T_l$ (resp. $T'_1,\dots,T'_{l'}$) the trees of depth at most $d$ rooted at the offsprings of the root of $N$ (resp. $N'$). The induction is found to be (see~\cite{piccioli_aligning_2022} for details)
\beq
L^{(d+1)}(N,N')= f(l,l';\{L^{(d)}(T_i,T'_{i'})\}_{i \in [l],i'\in[l']}) \ ,
\label{eq:theLformula}
\eeq
the function $f$ taking as input two integers $l,l'$ and an $l\times l'$ matrix. It is defined by
\begin{equation}
f(l,l';\{L_{i,i'}\})=
\underset{k=0}{\overset{\min(l,l')}{\sum}} e^{\lambda s}(1-s)^{l+l'}\left(\frac{s}{\lambda(1-s)^2}\right)^{k}\underset{I,I',\sigma}{\sum} \ \underset{i\in I}{\prod} L_{i,\sigma(i)} \ ,
\label{eq:alg-func}
\end{equation}
where $I$ (resp. $I'$) is a subset of $[l]$ (resp. of $[l']$) containing $k$ elements, $\sigma$ is a bijection from $I$ to $I'$, and where in the term $k=0$ the last sum is by convention equal to 1. The induction is initialized with the boundary condition for $d=0$, namely $L^{(0)}(T,T')=1$.

These considerations lead naturally to a message-passing algorithm for the alignment of random graphs: the quantities $L^{(d)}_{i,i'}$  that appear in (\ref{eq:trunc-post}) are computed as $L^{(d)}_{i,i'}=L^{(d)}(G_{i,d},G'_{i',d})$ in the thermodynamic limit, and interpreted as ``scores'' proportional to the probability that $i$ and $i'$ are matched in the hidden permutation. Hence, an estimator $\hpi$ can be constructed by choosing, for each $i$, the vertex $i'$ maximizing the score (recall the expression (\ref{eq:argmaxi}) of the optimal estimator with the exact posterior distribution). Moreover, the recursive nature of trees and of $L^{(d)}$ expressed in Eq.~(\ref{eq:theLformula}) allows to compute these quantities by passing messages between pairs of vertices. To define this procedure more explicitly we recall that $\partial i$ denotes the set of vertices that are adjacent to a vertex $i$ in $G$, similarly $\partial i'$ is the set of neighbors of $i'$ in $G'$, and $d_i = |\partial i|$, $d_{i'}=|\partial i'|$ are the corresponding degrees. We introduce messages $L^{(t)}_{ii'\to jj'}$, where $i i'$ runs over all pairs of vertices of $G$ and $G'$, $j\in\partial i$ and $j'\in \partial i'$, and $t \in \{0,1,\dots,d-1\}$ should be thought of as a discrete time index. The algorithm of~\cite{ganassali_correlation_2022,piccioli_aligning_2022} can then be summarized as follows:

\begin{enumerate}
\item initialize all messages to $L_{ii' \to jj'}^{(0)}=1$.
\item for each $t=1,\dots,d-1$, update them according to
\begin{equation}
L_{ii' \to jj'}^{(t)} = f(d_i-1,d_{i'}-1;\{L_{kk' \to ii'}^{(t-1)} \colon k \in \partial i \setminus j ,  k' \in \partial i' \setminus j' \} ) \ ,
\end{equation}
where the function $f$ has been defined in Eq.~(\ref{eq:alg-func}).
\item compute the scores between all pairs of vertices as
\begin{equation}
L_{i,i'}^{(d)} = f(d_i,d_{i'};\{L_{jj' \to ii'}^{(d-1)}  \colon j \in \partial i ,  j' \in \partial i' \} ) \ .
\end{equation}
\item finally the estimator $\hpi$ is computed as $\hpi(i)=\arg\max_{i'} L^{(d)}_{i,i'}$, with ties broken uniformly at random if several $i'$ achieve the same maximal score. Note that the inference of the vertex matched to $i$ is correct, i.e. $\hpi(i)=\pi_\star(i)$, if the vector $\{L^{(d)}_{i , \cdot}\}$ is maximal in $i'=\pi_\star(i)$.
\end{enumerate}
The numerical results of~\cite{piccioli_aligning_2022}, as well as the analytical considerations of~\cite{ganassali_correlation_2022,ganassali_statistical_2022}, show that this algorithm (with $d = \Theta(\ln n)$) succeeds in the partial recovery task in a sizable part of the parameter space $(\lambda,s)$ (and in particular for any $s>\sqrt{\alpha}$ when $\lambda$ is large enough, where $\alpha$ is Otter's constant). A drawback of this approach is however its relatively high computational cost: the recursion function $f$ of Eq.~(\ref{eq:alg-func}) contains a sum over the permutations of the offsprings of the roots (translating the ignorance of their true alignment), the number of terms to compute thus grows factorially with the degrees $l,l'$ of the considered vertices. Since in a random graph of size $n$ the maximum degree grows slowly with $n$ (as $O((\log n)/(\log\log n))$) the algorithm has formally a polynomial running time for all finite $\lambda$, but in practice this growth limits quite drastically the range of $\lambda$ that are accessible to numerical simulations. Our goal in the following is to present variations of this algorithm that have a reduced computational cost, at the price of a slightly deteriorated accuracy of estimation.

\subsection{A family of approximations}
\label{sec:approx-alg}

The new algorithms that we will propose in the present work are based on a result of~\cite{ganassali_statistical_2022} that provides an alternative expression of the likelihood ratio $L^{(d)}$, equivalent to the recursive formula (\ref{eq:theLformula}). We recall the notation $\chi_d$ for the set of unlabeled trees of depth at most $d$, to which the arguments $T,T'$ of $L^{(d)}$ belong. Theorem 4 in~\cite{ganassali_statistical_2022} states that the likelihood ratio can be ``diagonalized'' as
\begin{equation}
L^{(d)}(T,T')=\sum_{\beta\in\chi_d} s^{|\beta|-1} g_\beta^{(d)}(T) g_\beta^{(d)}(T') \ ,
\label{eq:theL2formula}
\end{equation}
where the ``eigenvectors'' $g_\beta^{(d)}$ are functions from $\chi_d$ to $\mathbb{R}$, indexed by ``dual'' trees $\beta \in \chi_d$, while the associated ``eigenvalues'' are $s^{|\beta|-1}$, with $|\beta|$ denoting the size of the tree $\beta$, defined as the number of its vertices. We call ``dual'' the trees $\beta$ because they serves as an index for the eigenvectors, in analogy with the wavevector $k$ that can be seen as a dual of the position $x$ in a decomposition into plane waves $e^{ikx}$. The explicit definition of $g_\beta^{(d)}(T)$ is rather technical and will be given in Appendix~\ref{app_ev_definitions}, a proof of the equivalence between the two definitions (\ref{eq:theLformula},\ref{eq:theL2formula}) of $L^{(d)}$ can be found in Appendix~\ref{app:proof-truncation}. Some explicit examples and further properties of the functions $g_\beta^{(d)}$ will also be presented in the Appendices~\ref{app_ev_definitions} and \ref{app:martingale}. Let us just mention at this point that the functions $g_\beta^{(d)}$ depend on $\lambda$ but not on $s$, and that they form a basis of orthogonal polynomials in the space $\chi_d$ with respect to the distribution $\pzero{d}{}$, in the sense that
\beq
\underset{T \in\chi_d}{\sum} \pzero{d}{T} g_\beta^{(d)}(T) g_{\beta'}^{(d)}(T) = \delta_{\beta,\beta'} \ .
\label{eq_orthogonality}
\eeq
In addition, the eigenvector corresponding to the trivial dual tree $\beta = \bullet$ containing only the root is the constant function, $g_\bullet^{(d)}(T)=1$; more generically the depth of a dual tree $\beta$ controls the number of generations of $T$ on which $g_\beta^{(d)}$ actually depends (see App.~\ref{app:martingale} for more details on this point).

Our approach will consist in approximating the function $L^{(d)}$ by including only a part of the terms in the summation of~(\ref{eq:theL2formula}), following the rough intuition that each term $\beta$ of the sum captures one piece of information on the trees $(T,T')$, as for instance the Fourier decomposition of a signal in various harmonics, hence any partial summation should discard a part of the information but might keep enough of it to still achieve our original inference goal. More concretely, for a given subfamily $\hchi_d \subset \chi_d$ of the unlabeled trees of depth at most $d$, let us denote
\begin{equation}
\hL^{(d)}(T,T')=\sum_{\beta\in\hchi_d} s^{|\beta|-1} g_\beta^{(d)}(T) g_\beta^{(d)}(T') \ .
\label{eq_def_hLd}
\end{equation}
In the following we will consider that the trees in $\hchi_d$ obey the additional constraint that each vertex has at most $m$ descendants, where $m \ge 2$ is an integer parameter. It turns out that for this choice of $\hchi_d$ the approximate function $\hL^{(d)}(T,T')$ can be computed in a recursive way. Decomposing as above $N \in \chi_{d+1}$ as a root vertex with $l$ descendants, themselves being the roots of the trees $T_1,\dots,T_l$ of depth a most $d$ (and a similar decomposition for $N'$), one finds an analog of Eq.~(\ref{eq:theLformula}),
\beq
\hL^{(d+1)}(N,N')= \hf(l,l';\{\hL^{(d)}(T_i,T'_{i'})\}_{i \in [l],i'\in[l']}) \ ,
\label{eq_hLd_recursion}
\eeq
where $\hf$ is the expansion in powers of $s$ of the function $f$ given in Eq.~(\ref{eq:alg-func}), up to order $s^m$. To be more explicit, let us introduce a convenient notation for the manipulation of formal power series: if $h(t)=\sum_{n \ge 0}h_n t^n $, one writes $[t^p]h(t) = h_p$ for the extraction of the coefficient multiplying the monomial $t^p$ in the series. With this convention the approximated recursion function $\hf$ reads:
\begin{equation}
\hf(l,l';\{\hL_{i,i'}\})=\sum_{p=0}^m s^p[t^p]e^{\lambda t}(1-t)^{l+l'}\sum_{k=0}^{\min(l,l')}\left(\frac{t}{\lambda(1-t)^2}\right)^k\sum_{I,I',\sigma}\prod_{i\in I}\hL_{i,\sigma(i)} \ ,
\label{eq:alg-func-trunc}
\end{equation}
where the summation over $I,I',\sigma$ has the same range as in Eq.~(\ref{eq:theLformula}); one can note that in the second sum only the terms with $k \le p$ do contribute.

The restriction of the dual trees $\beta$ summed over in (\ref{eq_def_hLd}) to those in which every vertex has at most $m$ offsprings thus translates into a truncation of the recursion function to terms of order at most $s^m$ in (\ref{eq:alg-func-trunc}); if this statement can be spelled out in a concise way, its proof is quite technical and relies on the precise definition of the eigenvectors $g_\beta^{(d)}$, for this reason we defer further details of its justification to the Appendix~\ref{app:proof-truncation}. Even if the notation does not make it explicit it should be clear that the functions $\hL^{(d)}$ and $\hf$ depend on the parameter $m$. The function defined in Eq.~(\ref{eq:alg-func-trunc}) can be easily implemented numerically for any value of $m$, by enumerating all subsets $I$ and $I'$ of $[l]$ and $[l']$ of cardinality $k$ (smaller than $m$); for a fixed $k$ the number of such subsets scales with the degrees as $O((ll')^k)$. It turns out that for small values of $m$ (and hence of $k$) one can reorganize these summations by exploiting some inclusion-exclusion principles in order to reduce this computational cost; the simplest non-trivial case $m=2$ yields
\begin{multline}
\hf(l,l';\{\hL_{i,i'}\})=  1 + s \left(\frac{1}{\lambda} \sum_{i,i'} \hL_{i,i'} + \lambda - l - l' \right) + \frac{1}{2} s^2
\left( \frac{1}{\lambda^2} \sum_{i,i'} (\hL_{i,i'})^2 - \frac{1}{\lambda^2} \sum_{i} \left(\sum_{i'} \hL_{i,i'}\right)^2 - \frac{1}{\lambda^2} \sum_{i'} \left(\sum_{i} \hL_{i,i'}\right)^2 \right. \\ \left. +\frac{4}{\lambda} \sum_{i,i'} \hL_{i,i'} - l - l' + \left(\frac{1}{\lambda} \sum_{i,i'} \hL_{i,i'} + \lambda - l - l' \right)^2   \right) \ ,
\label{eq:alg-func-m2}
\end{multline}
where the sums over $i$ (resp. $i'$) are from $1$ to $l$ (resp. $1$ to $l'$), and which can be evaluated with $O(ll')$ operations, smaller than the $O((ll')^2)$ cost of a naive computation from Eq.~(\ref{eq:alg-func-trunc}). The justification of (\ref{eq:alg-func-m2}) is presented in Appendix~\ref{app:general-approximation}, along with a similar, slightly longer formula for the case $m=3$ that can also be computed with $O(ll')$ operations, and some hints on how to generalize the computation to larger values of $m$. In the following we shall conventionally call $m=\infty$ the original approach presented in Sec.~\ref{sec:original-alg}, since it corresponds to remove the constraint on the structure of the dual trees, i.e. $\hL^{(d)} = L^{(d)}$ if $\hchi_d=\chi_d$ in Eq.~(\ref{eq_def_hLd}).

Coming back to the original graph alignment problem, the algorithmic strategy we propose is a variation of the original one recalled in Sec.~\ref{sec:original-alg}, where the scores $L_{i,i'}^{(d)}=L^{(d)}(G_{i,d},G'_{i',d})$ are replaced by their approximated counterparts, $\hL^{(d)}(G_{i,d},G'_{i',d})$. Since the latter still admits a recursive computation, the message-passing implementation of the algorithm has exactly the same structure, the only modification being the replacement of the recursion function $f$ of Eq.~(\ref{eq:alg-func}) by its approximation $\hf$ of Eq.~(\ref{eq:alg-func-trunc}) in the steps 2 and 3 of the algorithm. In the next Section we shall present the results of numerical simulations performed in this way; before that let us make a few remarks:
\begin{itemize}
\item a crucial advantage of the finite $m$ variation we just introduced is its much lower computational cost with respect to the original procedure. Recall indeed that the latter relied on the computation of the recursion function $f$, which incurs a factorial growth in the vertices' degrees of the number of operations required to perform the sums in Eq.~(\ref{eq:alg-func}). On the contrary the truncation in $\hf$ reduces this cost to a polynomial growth, with an exponent depending on $m$. For instance the $m=2$ formula displayed in Eq.~(\ref{eq:alg-func-m2}) can be computed in $O(ll')$ operations, and this is also the case for $m=3$ with an appropriate bookkeeping (see Appendix~\ref{app:general-approximation} for further details on the larger $m$ cases). As a consequence, a larger region of the $(\lambda,s)$ parameter space becomes accessible to numerical explorations. Since there are $n^2$ pairs of vertices to consider, over which the average of $l l'$ scales as $\lambda^2$ for large $\lambda$, this justifies the statement made in the introduction on the $O(n^2 (\log n)\lambda^2)$ runtime for $m \in \{2,3\}$ (the additional logarithmic factor coming from the scaling of the depth parameter $d$, as discussed below).
  
\item an important caveat is the somehow arbitrary and uncontrolled character of the approximation made in the finite $m$ computation. It is reminiscent of the so-called low-degree polynomial approach~\cite{kunisky_notes_2019} to hypothesis testing problem, since $\hL^{(d)}$ can be viewed as the orthogonal projection of $L^{(d)}$ onto a subspace of functions of $\chi_d \times \chi_d$ (where orthogonality is with respect to averages under $\pzero{d}{} \otimes \pzero{d}{}$ and follows from Eq.~(\ref{eq_orthogonality})). We emphasize however that this hypothesis task on trees (namely deciding whether the local neighborhoods of the two vertices $i$ and $i'$ are correlated or not) is only here a ``subroutine'' of our objective, i.e. the recovery of the hidden permutation between the graphs. The quality of the estimator for the latter task, as measured by the overlap (\ref{eq:overlap}) between $\pi_\star$ and $\hpi$, is thus not directly related to the accuracy of the approximation of $L^{(d)}$ by $\hL^{(d)}$ in a mean square error sense. As a consequence it is not a priori clear whether including more terms in the sum, i.e. increasing $m$, should yield a monotonic increase in the overlap. We will come back on this point with further analytical considerations in Sec. ~\ref{sec:trees-results}.

\item the interpretation of $L^{(d)}$ as a likelihood ratio (i.e. a ratio of probabilities) is no longer valid for its counterpart $\hL^{(d)}$, in particular the latter is no longer ensured to be positive, as the former was by definition.

\end{itemize}

\section{Numerical results}
\label{sec:numerical-results}

We present in this Section the results of numerical simulations of the algorithms introduced above; our main objective is to quantify the loss of accuracy incurred when replacing the original procedure of~\cite{piccioli_aligning_2022} (denoted $m=\infty$ in the following) by its finite $m$ variations (we will concentrate on $m=2$ and $m=3$, that are the two simplest non-trivial cases). As argued previously, the latter have much lower computational costs; we will not present systematic comparisons of their running times, since they are quite dependent on implementation details, but to give an example of the velocity gain let us mention that on a graph of $n=128$ vertices, with parameters $\lambda=3.0$, $s=0.86$, $d=20$, a simulation that took 94 seconds for the $m=\infty$ algorithm only lasted 11 seconds for the $m=2$ version, and 24 seconds for the $m=3$ one, which represents a notable acceleration. This gain is even more important for larger values of the average degree $\lambda$, the finite $m$ simulations can thus probe a regime of $\lambda$ that is not accessible in a reasonable time with the $m=\infty$ algorithm.

\subsection{The accuracy of the approximated scores}

\begin{figure}
\centering
\includegraphics[width=.49\textwidth]{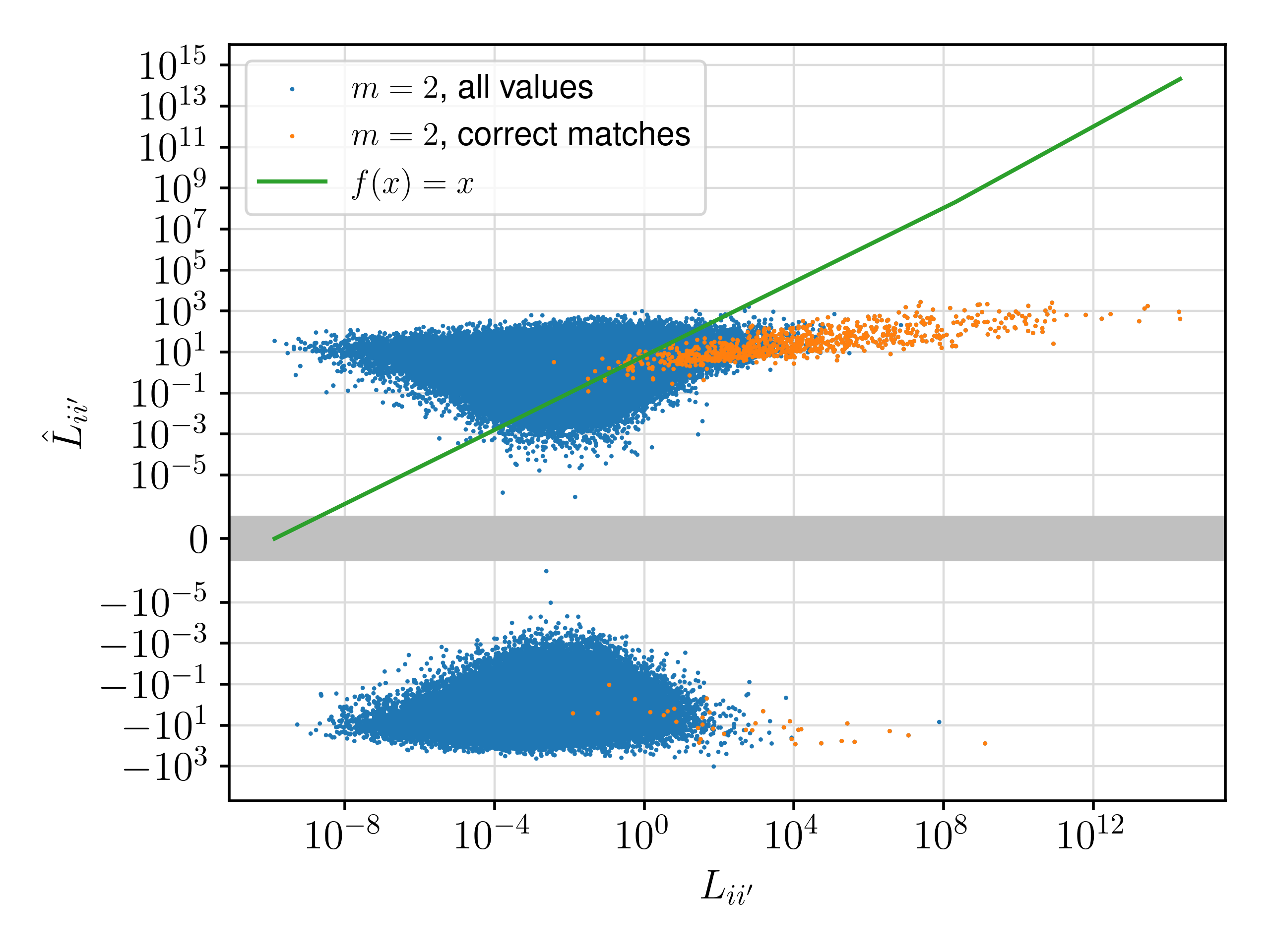}
\includegraphics[width=.49\textwidth]{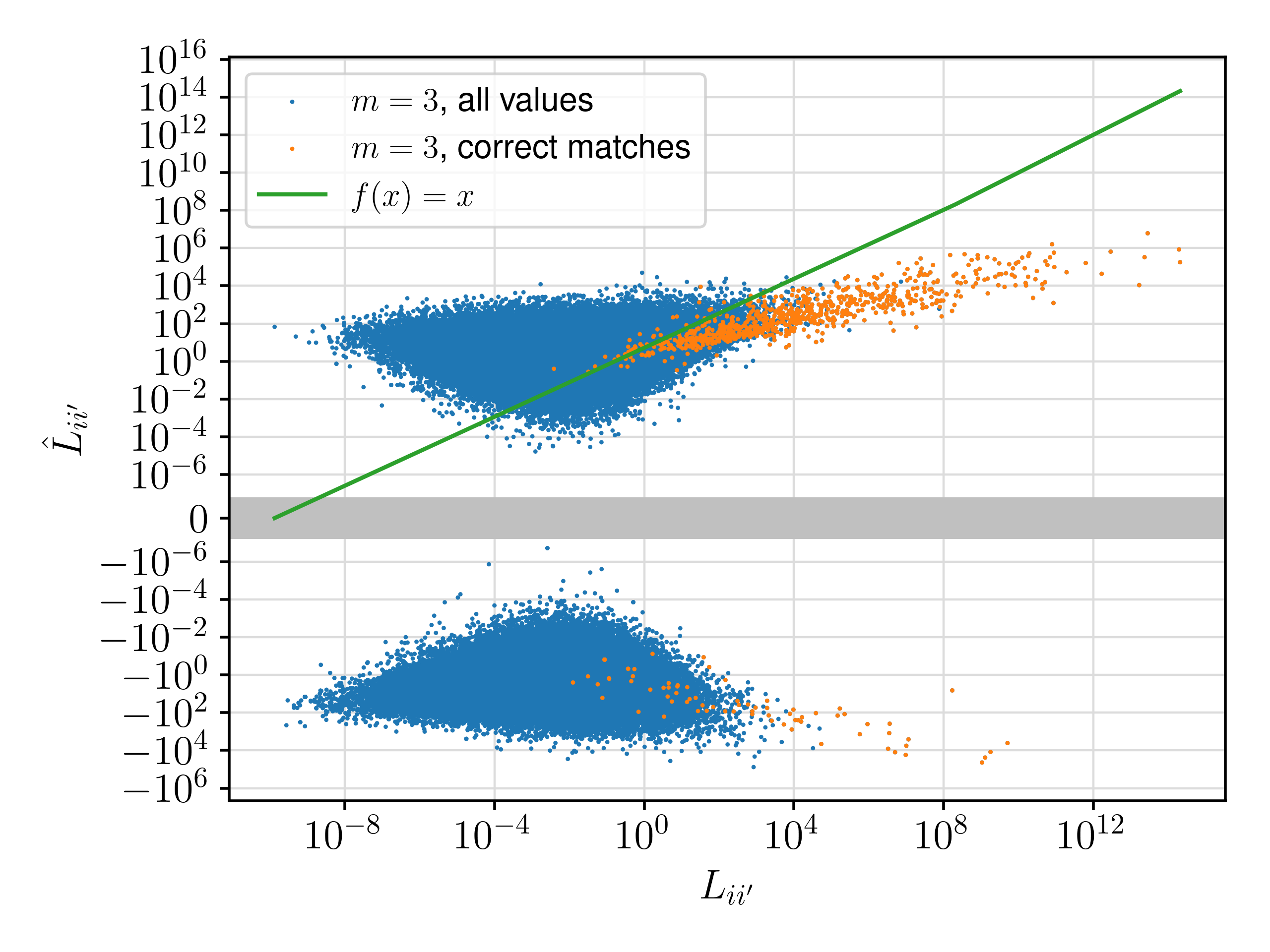}
\caption{Comparison of the scores $L_{i,i'}^{(d)}$ and their approximations $\hL_{i,i'}^{(d)}$ on a single graph with $n=1024$, $\lambda=2.4$, $s=0.9$, $d=4$, $m=2$ (left panel) and $m=3$ (right panel). Each point in these scatter plots corresponds to a pair $i,i'$ of vertices of both graphs, the horizontal axis corresponding to $L_{i,i'}^{(d)}$, the vertical one to $\hL_{i,i'}^{(d)}$. Correctly matched pairs with $i'=\pi_\star(i)$ are highlighted in red. The $x$ axis is in log-scale while the $y$ axis is logarithmic on both the positive and negative sides, with a linear region around the origin (in gray on the figure) to join them (we have shrunk it so that every dot displayed is actually in log-scale).}
\label{fig:LvsLhat-scatter}
\end{figure}

In a first type of experiment we compared, on a single graph of size $n$, the $n^2$ exact scores $L_{i,i'}^{(d)}$ with their approximated counterparts $\hL_{i,i'}^{(d)}$; these results are presented as scatter plots in Fig.~\ref{fig:LvsLhat-scatter}, for the parameters $n=1024$, $\lambda=2.4$, $s=0.9$, $d=4$, $m=2$ (left panel) and $m=3$ (right panel). From a first look at these figures one may conclude that the approximations are of a very bad quality, since the cloud of points strongly deviates from the diagonal (drawn as a green line) one would obtain if $\hL_{i,i'}^{(d)} = L_{i,i'}^{(d)}$. A moment of thought reveals however a more encouraging situation: as a matter of fact what is important for the correct functioning of the algorithm is not the value of the scores themselves, but rather the fact that they are larger for pairs of vertices corresponding to correct matches in the sought for permutation than for the other pairs. Recall indeed that the vertex $i$ will be correctly aligned by the algorithm if $\hpi(i)=\pi_\star(i)$, which happens if $\hL^{(d)}_{i,\pi_\star(i)} > \hL^{(d)}_{i,i'}$ for all $i' \neq \pi_\star(i)$. In Fig.~\ref{fig:LvsLhat-scatter} we have highlighted in red the points corresponding to the correctly matched pairs of vertices ($i' = \pi_\star(i)$): for these there is a clear positive correlation between $L_{i,i'}^{(d)}$ and $\hL_{i,i'}^{(d)}$, more marked for $m=3$ than for $m=2$, which suggests that the approximate algorithms may be able to pick the correct pairs of vertices among the much more numerous incorrect ones, even if $\hL_{i,i'}^{(d)}$ strongly differs from $L_{i,i'}^{(d)}$ for the latter pairs. Fig.~\ref{fig:LvsLhat-scatter} also shows that $\hL_{i,i'}^{(d)}$ can be negative, at variance with $L_{i,i'}^{(d)}$, as anticipated above. Around half of the $\hL_{i,i'}^{(d)}$ are indeed negative, but this is the case of less than 10\% for the pairs of correctly matched vertices. 

Let us emphasize that the interpretation of these plots is delicate, since the scores are auxiliary quantities that are only used as an intermediate in the computation of the relevant object, namely the estimated vertex correspondence $\hpi$; moreover the scatter plots superpose the vectors $\hL_{i,\cdot}^{(d)}$ and $L_{i,\cdot}^{(d)}$ for all $i$'s, whereas the correct assignment of $\hpi(i)$ should be read from the relative ranking of $\pi_\star(i)$ in these vectors for a single $i$. Since we did not find a way of representing these rankings that would be visually easy to read we presented these scatterplots that give at least an idea of the behavior of the approximated scores. In the following we concentrate on the central quantity of our study, namely the estimator $\hpi$ of the hidden permutation and its quality measured as its overlap with $\pi_\star$.

\subsection{The dependency on the depth parameter}

\begin{figure}
\centering
\includegraphics[width=.49\textwidth]{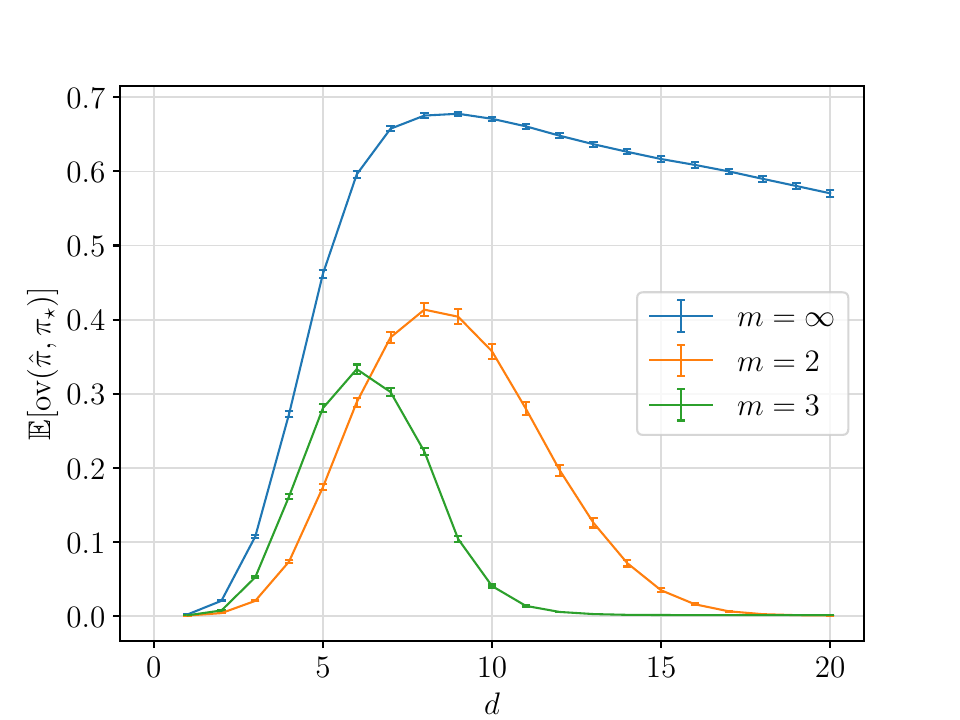}
\includegraphics[width=.49\textwidth]{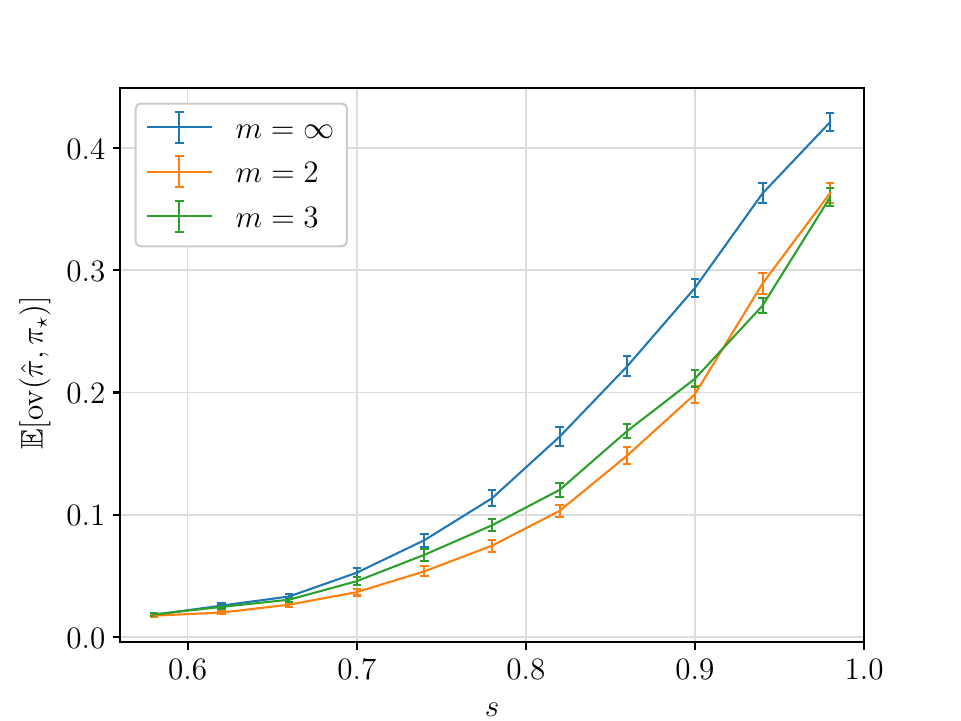}
\caption{Left panel: the average overlap between the true and estimated permutations as a function of the depth parameter $d$ for $n=2048$, $\lambda=2.4$, $s=0.86$, $m=\infty$ (blue curve), $m=2$ (orange curve) and $m=3$ (green curve). Each point corresponds to an average over 50 independent samples.
Right panel: The average overlap as a function of $s$ for $n=512$, $\lambda=1.2$, $m=\infty$ (blue curve), $m=2$ (orange curve) and $m=3$ (green curve). Each point is an average over 50 samples.
}
\label{fig:overlap-depth_and_overlap-full-vs-approx}
\end{figure}

The complete specification of the algorithmic procedure for the graph alignment requires a choice of the parameter $d$, that controls the depth of the neighborhoods of the vertices $i$ and $i'$ explored in order to decide whether $i$ is matched to $i'$ or not. A priori larger values of $d$ should lead to better alignments, since they exploit a larger amount of information contained in the graphs; however for a fixed size $n$ the locally tree-like character of the random graphs, on which our analysis crucially relies, is no longer valid when $d$ exceeds some multiple of $\log n$. As a consequence a compromise should be found between these two effects, and one can expect the existence of an optimal depth parameter $d_\star$, depending on all the other parameters of the problem ($n$, $\lambda$, $s$ and $m$), at which the overlap between the true permutation $\pi_\star$ and its estimate $\hpi$ is maximal. This was indeed observed in~\cite{piccioli_aligning_2022} for the $m=\infty$ version of the algorithm, and we show in the left panel of Fig.~\ref{fig:overlap-depth_and_overlap-full-vs-approx} that this phenomenon also occurs for $m=2$ and $m=3$.

The results presented in the rest of this Section have been obtained with the parameter $d$ adjusted to its optimal value $d_\star$; in practice we repeated the simulations with $d$ in the range $[1,20]$, and chose the one maximizing the overlap for each considered value of $n$, $\lambda$, $s$ and $m$. The optimal parameter $d_\star$ has to grow (logarithmically) with $n$: for the same reasons that were discussed in~\cite{piccioli_aligning_2022} in the $m=\infty$ case, partial recovery cannot be achieved if the large $n$ limit is taken with $d$ fixed.

\subsection{The overlap reached by the approximate algorithms}

We shall now present a series of results on the accuracy of the graph alignment produced by the approximate algorithm, measured in terms of the average overlap between the reconstructed permutation and the true one, as a function of the parameters $n$, $\lambda$, $s$ and $m$.

In the right panel of Fig.~\ref{fig:overlap-depth_and_overlap-full-vs-approx} we plot this average overlap as a function of the correlation parameter $s$, for relatively small size $n$ and average degree $\lambda$, in such a way that the $m=\infty$ algorithm can be run in a reasonable time and thus compared to the $m=2$ and $m=3$ approximations. One can see on this figure that the overlap is slightly lowered by the approximation, but the range of $s$ in which a positive fraction of the hidden permutation is recovered remains roughly the same.

\begin{figure}
\centering
\includegraphics[width=.49\textwidth]{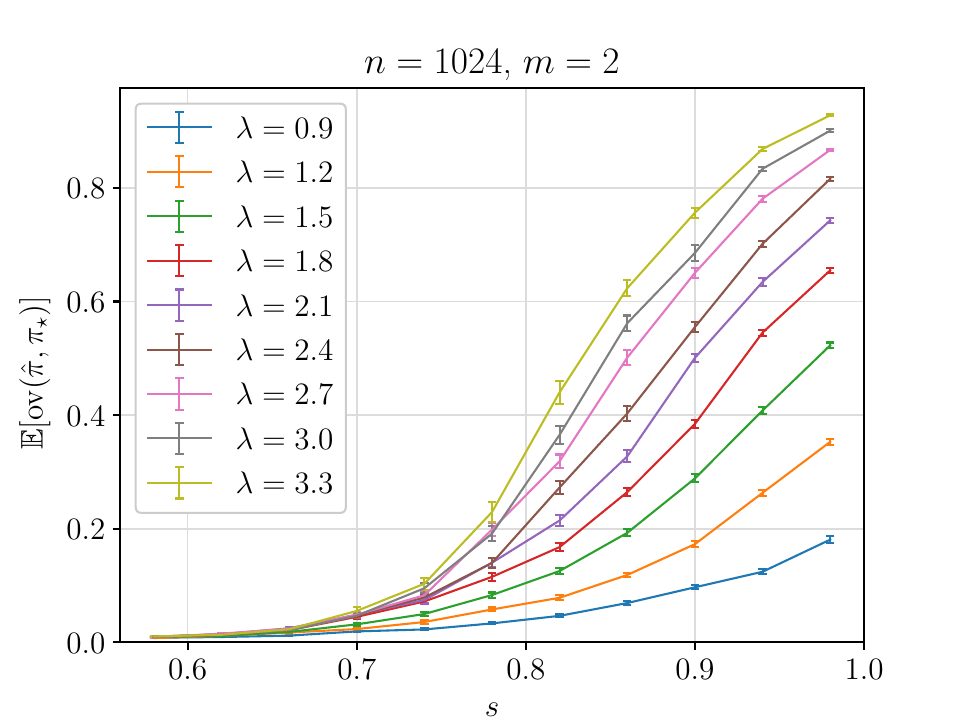}
\includegraphics[width=.49\textwidth]{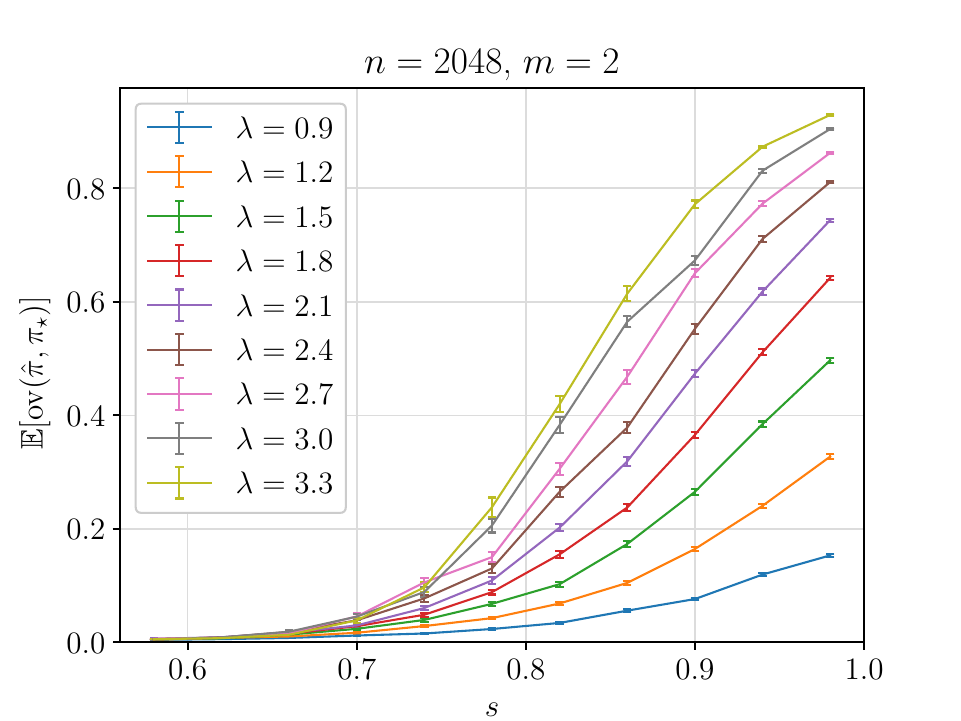}

\includegraphics[width=.49\textwidth]{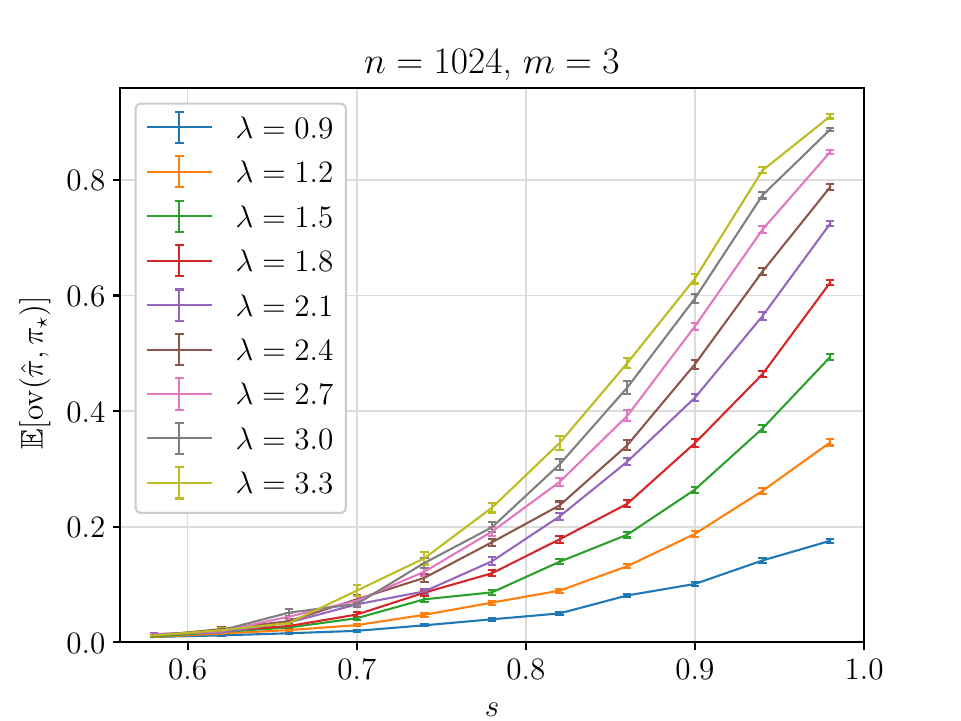}
\includegraphics[width=.49\textwidth]{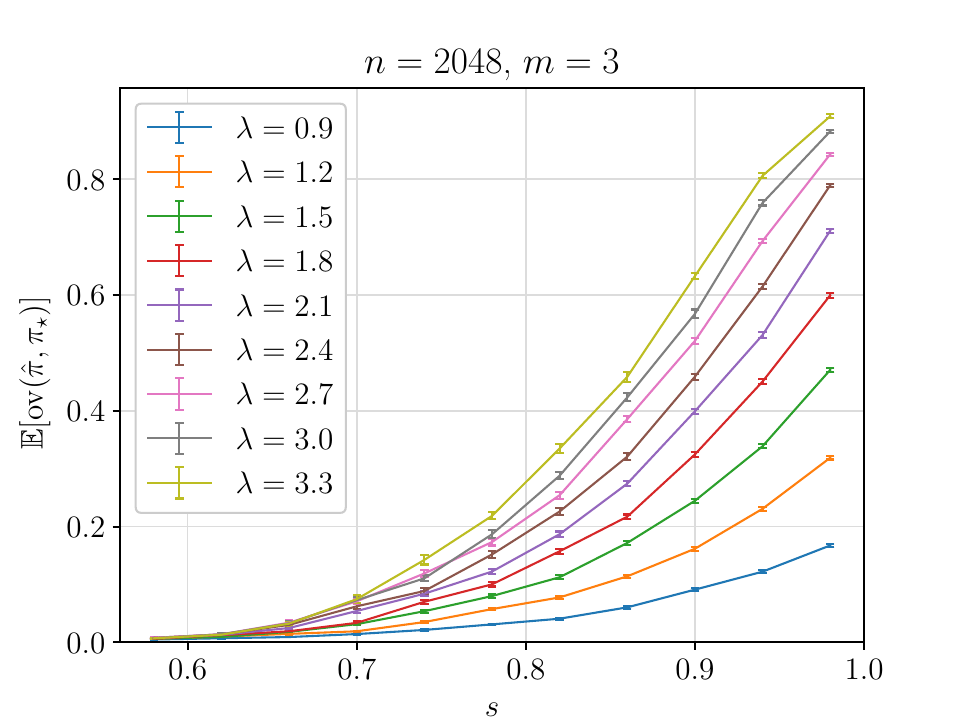}
\caption{The average overlap as a function of $s$, for $n=1024$ (left column) and $n=2048$ (right column), $m=2$ (top row) and $m=3$ (bottom row), and several values of $\lambda$ (see the keys). Each point is an average over 50 samples.}
\label{fig:overlaps}
\end{figure}

In Fig.~\ref{fig:overlaps} we report similar curves for larger sizes and average degrees, which are not all accessible to the $m=\infty$ algorithm because of computational cost issues. One can notice that all these curves increase monotonically with $s$, and become sharper when $n$ and $\lambda$ get larger (as was observed for $m=\infty$ in~\cite{piccioli_aligning_2022}).

\begin{figure}
\centering
\includegraphics[width=.49\textwidth]{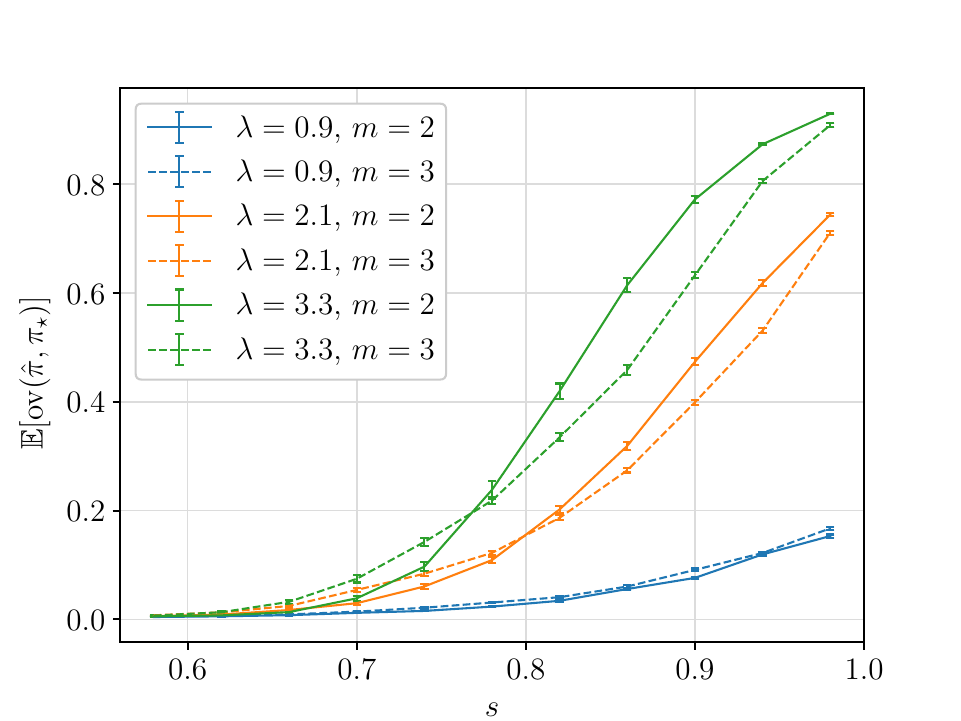}
\includegraphics[width=.49\textwidth]{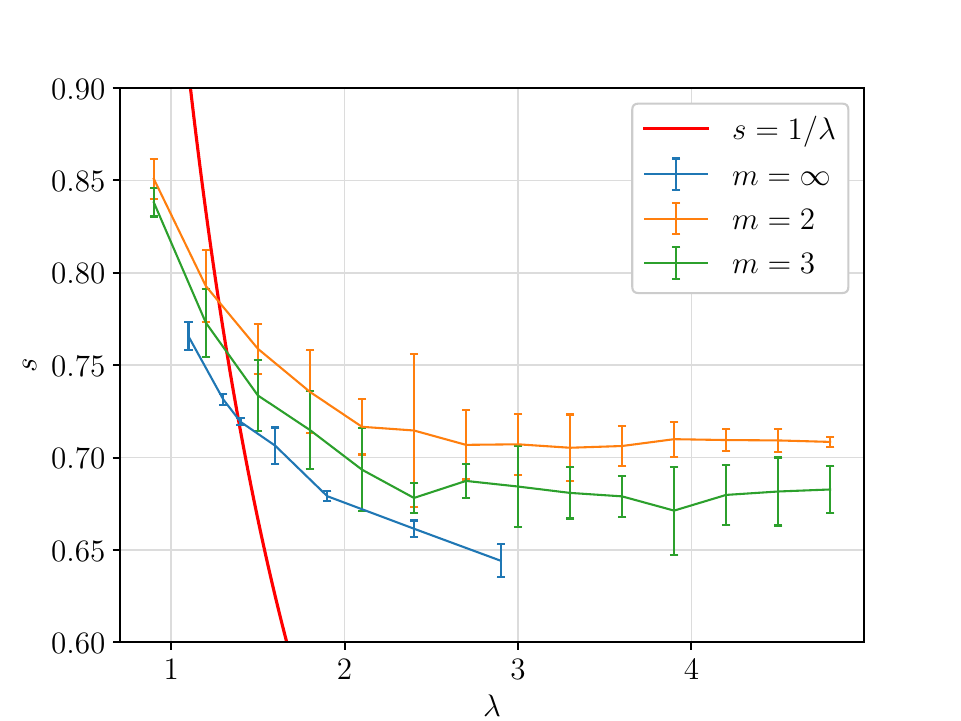}
\caption{Left panel: a comparison of the average overlaps achieved by the $m=2$ and $m=3$ versions of the algorithm, for $n=2048$ and three different values of $\lambda$; the data are the same as in the top right and bottom right panels of Fig.~\ref{fig:overlaps}.
Right panel: the effective phase diagram in the $(\lambda,s)$ plane, the transition lines being defined as the minimal value of $s$ such that the average overlap becomes larger than $R=0.05$, on graphs of size $n=2048$. The $m=\infty$ data is taken from~\cite{piccioli_aligning_2022}, the red line corresponds to the information-theoretic threshold $\lambda s=1$.}
\label{fig:overlaps-2vs3_and_phase-diag-approx-vs-full}
\end{figure}

A more systematic comparison of the $m=2$ and $m=3$ results is proposed in the left panel of Fig.~\ref{fig:overlaps-2vs3_and_phase-diag-approx-vs-full}, for a few different values of $\lambda$. One can see on this plot that, for a given value of $n$ and $\lambda$, the $m=2$ and $m=3$ overlap curves cross each other as a function of $s$ (this was also observable in Fig.~\ref{fig:overlap-depth_and_overlap-full-vs-approx} for other values of $n$ and $\lambda$), in other words for some parameters the $m=3$ algorithm performs worse than the $m=2$ one, which may sound rather counterintuitive. As a matter of fact it was mentioned at the end of Sec.~\ref{sec:approx-alg} that the quality of the approximation had no reason to be a monotonically increasing function of $m$, since the inclusion of more terms in the summation of Eq.~(\ref{eq_def_hLd}) does not guarantee per se a better accuracy of the estimation. It is nevertheless somehow comforting that for intermediate values of $s$ the $m=3$ variant reaches a notably better overlap than the $m=2$ one (for instance for $\lambda=3.3$, $s=0.65$), since this means that the domain of the $(\lambda,s)$ parameter space in which partial recovery is achieved grows with $m$, as discussed in the following.

\subsection{Phase diagram}

One can expect that in the thermodynamic limit the algorithms undergo a phase transition at a ($m$-dependent) critical value $\widehat{s}_{\rm algo}(\lambda)$ of the correlation parameter, in the sense that they achieve a partial recovery of the hidden permutation $\pi_\star$ if and only if $s> \widehat{s}_{\rm algo}(\lambda)$. To determine this threshold one should extrapolate the values of $\mathbb{E}[\ov(\hpi,\pi_\star)]$ obtained numerically at finite $n$ in order to determine their $n\to \infty$ limits, which by definition of  $\widehat{s}_{\rm algo}$ vanish for $s<\widehat{s}_{\rm algo}$ and are strictly positive for $s > \widehat{s}_{\rm algo}$. This extrapolation is unfortunately a very challenging numerical problem, since the finite-size effects are rather strong and since we do not have compelling analytical arguments on the form of these corrections that could guide the numerical fits. We therefore settled for a less ambitious procedure: from the overlap curves obtained at the largest size that we could reach (i.e. $n=2048$) we defined an effective threshold for the algorithmic transition (or more precisely crossover) as the minimal correlation parameter $s$ such that the average overlap grows above an arbitrarily chosen small value $R$ (we took $R=0.05$ for the results presented below). In practice we performed a quadratic fit of the overlap curves in the range of $s$ around the crossover, and computed the value of $s$ for which the fitted function crossed $R$. The outcomes of this procedure are displayed in the right panel of Fig.~\ref{fig:overlaps-2vs3_and_phase-diag-approx-vs-full} for $m=2$ and $m=3$, along with the $m=\infty$ result of~\cite{piccioli_aligning_2022} (for the same values of $R$ and $n$). We note that these transition lines enter the regime $\lambda s <1$ (see the red curve in the figure) where the information-theoretic results of~\cite{pmlr-v134-ganassali21a} forbids the possibility of partial recovery: this is a consequence of the finite-size corrections, and would disappear in the $n \to \infty$ limit. These results show that the domain of parameters in the $(\lambda,s)$ plane in which the algorithms achieve a partial recovery of the hidden permutation grows with $m$; moreover the transition line of the $m=3$ algorithm is quantitatively quite close to the $m=\infty$ one, which is rather satisfactory given the huge reduction in computational complexity provided by this approximation. The lines of transition for $m=2$ and $m=3$ seem to reach a finite value in the large $\lambda$ limit; we shall come back to this point in Sec.~\ref{sec:trees-results}, and propose a conjecture for these limits.

\subsection{The inefficiency of the naive truncation}

\begin{figure}
\centering
\includegraphics[width=.65\textwidth]{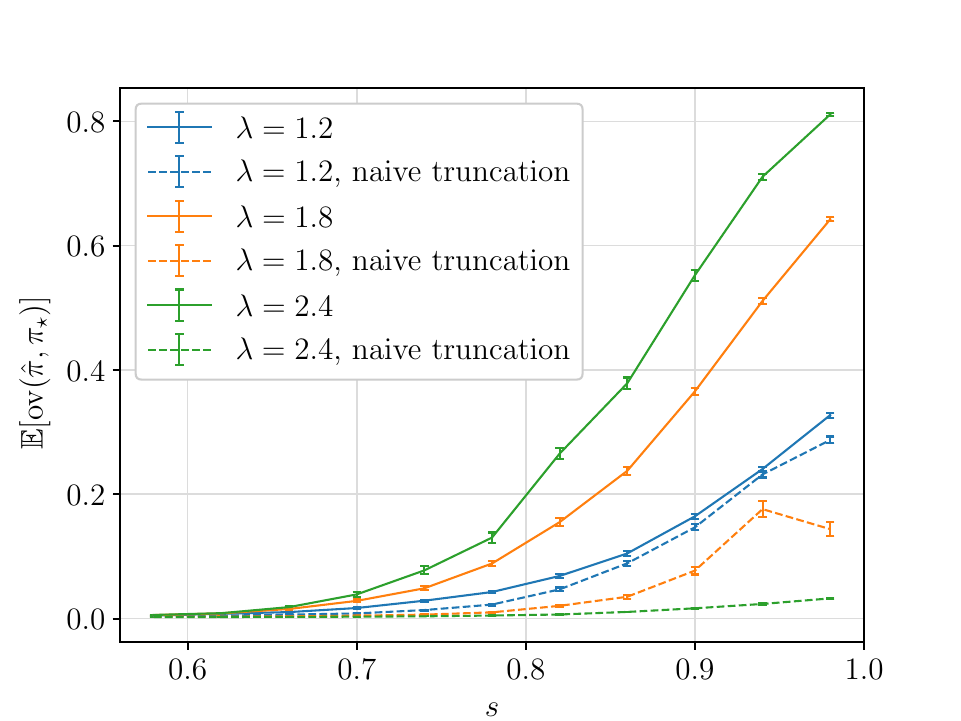}
\caption{The comparison of the average overlap reached by the $m=2$ algorithm (solid lines) and the naive truncation of Eqs.~(\ref{eq:function-tf},\ref{eq:another-coeff}) (dashed lines) for three values of $\lambda$, each point being averaged on 50 samples of graphs of size $n=2048$.}
\label{fig:another-overlap}
\end{figure}

We mentioned previously that the derivation of the finite $m$ version of the algorithm was not justified by a systematic optimization of the average overlap among a variational family of functions $\hL^{(d)}(T,T')$ used to compute the approximated scores. One can thus wonder if better accuracies could be reached with other functions that share the reduced computational cost of our approach. A natural family of such functions are those that can be computed in a recursive way as in (\ref{eq:theLformula}), with an ad-hoc choice of the recursion function $f$. If one restrict to functions $f$ that are polynomials of degree $m$ in the entries of the array $L$, and that respect the permutation symmetry between the subtrees, one is led to consider
\begin{equation}
\tf(l,l';\{L_{i,i'}\})= \sum_{k=0}^{\min(m,l,l')} C(k,l,l')  \sum_{I,I',\sigma} \prod_{i \in I} L_{i,\sigma(i)} \ , 
\label{eq:function-tf}
\end{equation}
where as before $I$ and $I'$ are subsets of $[l]$ and $[l']$ of $k$ elements, $\sigma$ is a bijection from $I$ to $I'$, and the $C(k,l,l')$ are now a priori arbitrary coefficients (that should also depend on the parameters $m$, $\lambda$ and $s$, and could possibly depend on the depth $d$). It would thus be impossible to explore systematically all the possible choices of these parameters and look for the ones that yield the best reconstruction accuracy with numerical simulations. Moreover the connection between the parameters $C$ and the average overlap is very indirect, we do not have an analytic formula expressing the latter in terms of the former (even in the case of the original algorithm at $m=\infty$, see Sec.~\ref{sec:trees-results} for further discussions of this difficulty), hence no starting point for a systematic optimization of the parameters in this family of algorithms. We therefore only considered the most naive choice, which amounts to truncate the sum in Eq.~(\ref{eq:alg-func}) without expanding it in powers of $s$, and thus yields
\begin{equation}
C(k,l,l') =e^{\lambda s} (1-s)^{l+l'} \left( \frac{s}{\lambda(1-s)^2} \right)^k \ .
\label{eq:another-coeff}
\end{equation}
The results of these simulations are presented in Fig.~\ref{fig:another-overlap}, where we compare, for $m=2$, the average overlap achieved with this naive truncation and the one based on Eq.~(\ref{eq:alg-func-m2}). One can see that when $\lambda$ grows the naive truncation performs very poorly (when $\lambda$ is much smaller than $m$ the discarded coefficients are negligible and the two approximations have similar accuracies); even if this does not prove that the choice made in Eq.~(\ref{eq:alg-func-m2}) is the best among the polynomially recursive score evaluation of degree 2, it shows at least that its performances cannot be reproduced by a naive truncation strategy.

\subsection{Comparison to other algorithms}

We present in this section the results of numerical simulations that confront the performances of our proposal to those of other algorithms. To the best of our knowledge the only other polynomial-time algorithm with performance guarantees for the partial recovery task in the sparse regime (with degrees kept finite in the large size limit), that does not fall into the family of message-passing algorithms discussed here, is the one of~\cite{mao_random_2023} (the information-theoretical optimal one of~\cite{ding_matching_2022} does not run in polynomial time). More precisely, the authors of~\cite{mao_random_2023} have shown that for all $s > \sqrt{\alpha}$ and any $o<1$ there exists a $\lambda_0(s,o)$ such that for all $\lambda \ge \lambda_0$ their algorithm performs the partial recovery task with an average overlap at least $o$, with a computational complexity $O(n^c)$ where the exponent $c$ only depends on $s$ (this paper contains also results in the regime of degrees growing with the system size). Unfortunately their algorithm seems more thought as a conceptual tool for proof than in view of a concrete implementation; it relies indeed on various subtasks, enumeration of a family of trees with the procedure of~\cite{BeHe80}, computation of the number of their automorphisms~\cite{CoBo81}, color coding to approximate a counting task~\cite{AlYuZw95,mao_testing_2022}, that are by themselves complicated problems with pseudocodes that do not seem to have been converted in actual codes in the previous litterature. An additional difficulty in the implementation of this algorithm relies in the actual choice of several constants, that are shown to exist by the rigorous proof but whose adequate values are not obvious a priori. 

\begin{figure}
\centering
\includegraphics[width=.65\textwidth]{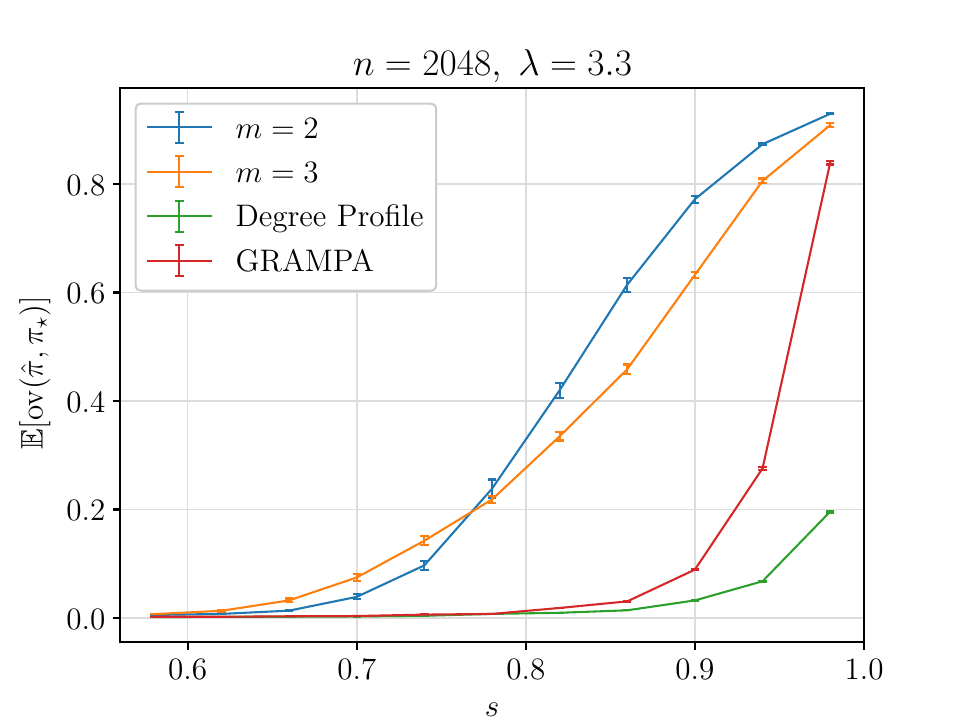}
\caption{The average overlap reached by the $m=2$ algorithm (blue), the $m=3$ version (orange), the degree profile algorithm of~\cite{ding_efficient_2018} (green), and the spectral one of~\cite{fan_spectral_2019,fan_spectral_2019_2} (red), for graphs of size $n=2048$ and average degree $\lambda=3.3$, as a function of the correlation parameter $s$. The first two lines share the data of some curves of Fig.~\ref{fig:overlaps}, the last two are obtained averaging over 10 samples for each point. For the GRAMPA algorithm the parameter $\eta$ was set to $0.2$, following the practice of~\cite{fan_spectral_2019,fan_spectral_2019_2}.
}
\label{fig:compa-algos}
\end{figure}

\begin{figure}
\centering
\includegraphics[width=.49\textwidth]{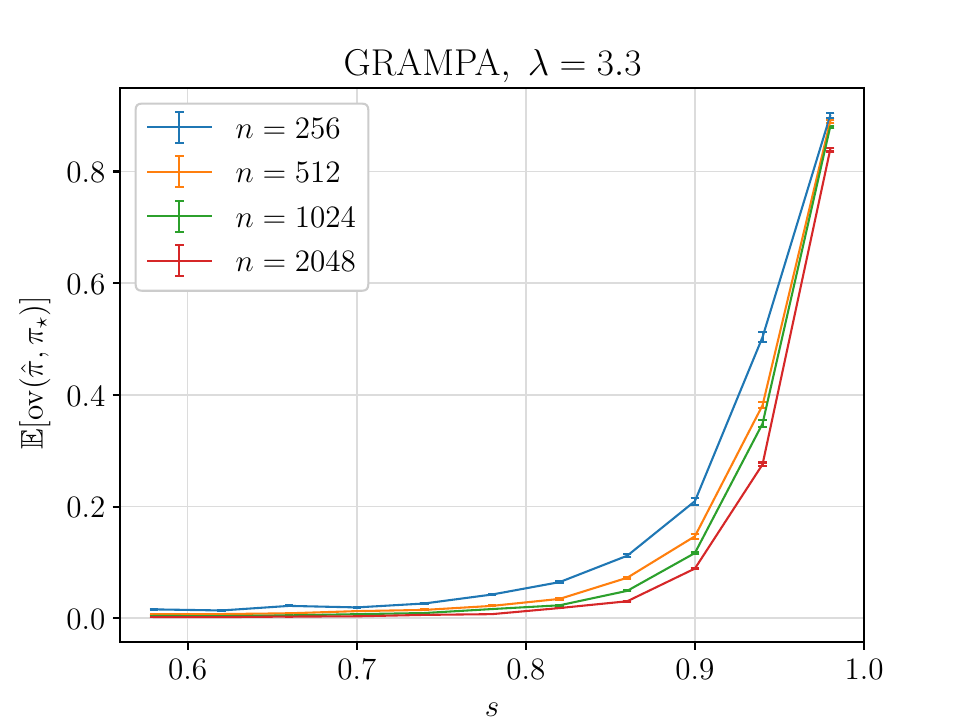}
\includegraphics[width=.49\textwidth]{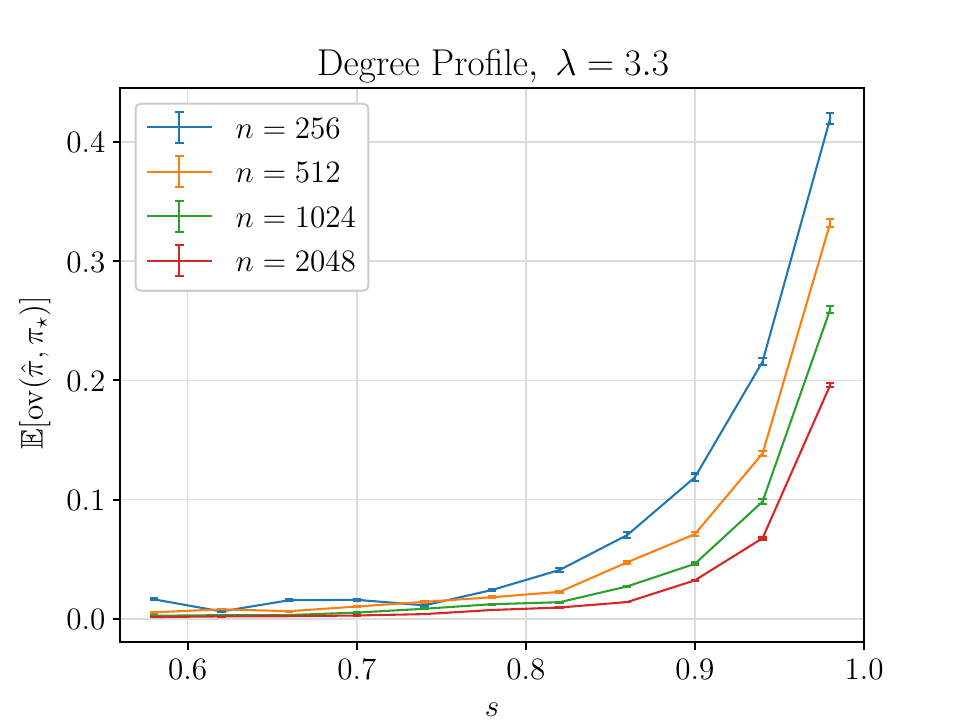}
\caption{The average overlap of the GRAMPA algorithm (left panel) and of the degree profile one (right panel) as a function of the correlation parameter $s$, for graphs of average degree $\lambda=3.3$ and various sizes. The points are averaged over 10 independent samples.}
\label{fig:compa-algos-nvarying}
\end{figure}

For these reasons we did not embark into the implementation of the algorithm of~\cite{mao_random_2023} (which was neither presented by their authors) and turned instead to the GRAMPA (for GRAph Matching by Pairwise eigen-Alignments) spectral algorithm of~\cite{fan_spectral_2019,fan_spectral_2019_2}, and the degree profile one of~\cite{ding_efficient_2018}. Both rely on the computation of similarity scores between vertices of the graphs to be aligned, the signature of a vertex being based on the eigendecomposition of the adjacency matrix for~\cite{fan_spectral_2019,fan_spectral_2019_2}, and on the statistics of the degrees of its neighbors for~\cite{ding_efficient_2018}. We present in Fig.~\ref{fig:compa-algos} the results of numerical simulations of these two algorithms (we used the implementation provided by the authors in~\cite{fan_spectral_2019_2_implementation,ding_efficient_2018_implementation}).
One can see that the spectral and degree profile algorithms require a larger level of correlation $s$ in order to reach sizable values of the overlap, and that they perform worse than our proposals, unless $s$ is very close to 1 in which case the performance of GRAMPA becomes comparable to ours (the improved quality of GRAMPA with respect to the degree profile algorithm was also seen in the numerical experiments of~\cite{fan_spectral_2019}). It is fair to underline that these two algorithms were actually devised for the exact recovery task, in the diverging degree regime, for which rigorous performance guarantees were proven in~\cite{ding_efficient_2018,fan_spectral_2019,fan_spectral_2019_2}. With this remark in mind one can find their performances reported in Fig.~\ref{fig:compa-algos} already quite satisfying, since they are obtained in a regime distinct from their natural objective (for $n$ finite the distinction between constant and diverging degrees is of course blurred and becomes a smooth crossover). We have checked that they degradates if $n$ increases with $\lambda$ kept fixed, see the results of Fig.~\ref{fig:compa-algos-nvarying}; as a matter of fact one can easily prove that the degree profile algorithm is bound to fail asymptotically in this regime, since there exists an extensive number of vertices that share the same signature and hence cannot be correctly aligned by this algorithm. This trend is quite clear on the right panel of Fig.~\ref{fig:compa-algos-nvarying}, while the decrease of the overlap of the GRAMPA algorithm (left panel) with the system size seems much slower.

\section{Analytical results and conjectures}
\label{sec:trees-results}

We shall present in this Section some analytical considerations on the expected behavior, in the thermodynamic limit $n \to \infty$, of the algorithms defined and studied numerically above. Several graph inference problems have been previously studied with statistical mechanics methods~\cite{zdeborova_statistical_2016}, for instance the Stochastic Block Model (SBM)~\cite{DeKrMoZd11,Mo17,Ab18}; in many cases it is possible to exploit the local convergence of random graphs towards trees in order to unveil quantitative connections between the large size limit of a graph problem and a related problem defined directly on trees. For the SBM example the limit object turns out to be the tree reconstruction problem~\cite{MoNeSl16}. In the present graph alignment case the connection is somehow subtler for several reasons: the limit involves pairs of trees instead of single ones, and the message passing algorithm to be studied is not obtained from a Belief Propagation approximation of a factor graph model (and its messages are passed among pairs of vertices and not along the edges of a single graph as is usually the case). 

Keeping these additional difficulties in mind let us start our discussion by considering the following question: when does a score-based graph alignment algorithm, i.e. one in which the estimator $\hpi$ is built as $\hpi(i)=\arg\max_{i'} S_{i,i'}$ for some scores $S$ between pairs of vertices of the two graphs, succeeds in the partial recovery task? By definition this occurs when, with a probability that remains strictly positive in the thermodynamic limit, $S_{i,\pi_\star(i)} > S_{i,i'}$ for all $i' \neq \pi_\star(i)$. Suppose that the scores $S_{i,i'}$ are computed from the local neighborhoods at distance at most $d$ of these vertices, i.e. are of the form $S^{(d)}(G_{i,d},G'_{i',d})$, and that $d$ grows sufficiently slowly with $n$ for these neighborhoods to be, with high probability, trees. Then $S_{i,\pi_\star(i)}$ is a random variable with the same distribution as $S^{(d)}(T,T')$ when $(T,T')$ is drawn from the correlated law $\pone{d}{T,T'}$; on the other hand the $S_{i,i'}$ with $i' \neq \pi_\star(i)$ are $n-1$ random variables distributed as $S^{(d)}(T,T')$, with $(T,T')$ drawn with probability $\pzero{d}{T} \pzero{d}{T'}$ (neglecting some correlations when $i'$ is at a distance smaller than $2d$ from $\pi_\star(i)$ in $G'$). The intuition provided by this reasoning is that partial recovery will be achieved if the ``typical'' values of $S^{(d)}$ under the law $\pone{d}{}$ are ``much larger'' than under $\pzero{d}{} \otimes \pzero{d}{}$, so that the unique representant $S_{i,\pi_\star(i)}$ of the former random variable has a chance to be larger than the $n-1$ copies $S_{i,i'}$ with $i' \neq \pi_\star(i)$ of the latter. Of course the expressions written in quotation marks above need to be defined more precisely, as we shall partially do in the following.

This discussion unveils a connection with hypothesis testing problems, a classical topic in statistics~\cite{CoTh_book}. In general these are formulated (in their simplest version) as follows: an observer is provided with a sample of a random object $X$, and has to decide whether $X$ was drawn from a law $\mathbb{P}_0$ (the null hypothesis), or from another law $\mathbb{P}_1$ (the alternative). The observer returns an estimate $t(X)\in \{0,1\}$ of the index of the law that generated $X$, whose quality can be quantified by the probability of the two types of errors (false positives and false negatives), $\gamma_0 = \mathbb{P}_0(t(X)=1)$ and $\gamma_1 = \mathbb{P}_1(t(X)=0)$. In usual hypothesis testing problems one considers the possibility of weak detection, where a test can outperform a random guess, or strong detection, where one requires the vanishing of both types of errors, asymptotically in some parameter of the problem. The tree hypothesis testing problem that appeared in the previous paragraph corresponds to distinguish whether $(T,T')$ has been drawn from $\pone{d}{}$ or from $\pzero{d}{} \otimes \pzero{d}{}$; the relevant question in this context is however not the possibility of weak or strong detection, but rather of one-sided detection, in which one requires that $\gamma_0$ vanishes as $d \to \infty$, while $\gamma_1$ remains strictly below $1$ in this limit. It is indeed in this regime that one will be able to pick the unique correct pair of vertices among the $n-1$ confounding ones, and it has been shown in~\cite{ganassali_correlation_2022} that an algorithm for the graph alignment problem that achieves the partial recovery of the hidden permutation can be built from a tree hypothesis testing procedure that performs one-sided detection.

Consider now the $m=\infty$ version of the algorithm, in which the score function $S^{(d)}$ coincides with the likelihood ratio $L^{(d)}$ defined in Eq.~(\ref{eq_def_L}). The general theory of hypothesis testing problems, and in particular the Neyman-Pearson lemma~\cite{neyman_problem_1933}, teaches that tests based on a thresholding of the likelihood ratio are optimal in the information-theoretic sense of minimizing one type of error while keeping the other one fixed to a constant. In a first (naive) attempt to quantify the difference of behavior of $L^{(d)}$ under the two laws of generation of the pair of trees, and hence its distinguishing power, one can compute its averages under the two laws. Let us therefore introduce some notations: for an arbitrary function $h(T,T')$ defined on $\chi_d \times \chi_d$ let us define the averages $\E_0^{(d)}[h]$ and $\E_1^{(d)}[h]$ as
\beq
\E_0^{(d)}[h] = \sum_{T,T'\in \chi_d} \pzero{d}{T} \pzero{d}{T'} \, h(T,T') \ , \qquad 
\E_1^{(d)}[h] = \sum_{T,T'\in \chi_d} \pone{d}{T,T'} \, h(T,T')  \ .
\eeq
Note that by definition of the likelihood ratio these two types of averages are related according to $\E_1^{(d)}[h] = \E_0^{(d)}[L^{(d)} h]$. In particular one has $\E_0^{(d)}[ L^{(d)}] = 1$ (applying the previous identity with $h=1$), which fixes the average value of the scores for independent trees. One can compute the corresponding quantity for the correlated law by exploiting the diagonalization result presented in Eq.~(\ref{eq:theL2formula}):
\begin{align}
    \E_1^{(d)}[ L^{(d)} ]& =\E_0^{(d)}[(L^{(d)})^2] \nonumber \\
    & = \sum_{\beta,\beta'\in\chi_d}s^{|\beta|+|\beta'|-2} \, \E_0^{(d)} \left[g_\beta^{(d)}(T)g_\beta^{(d)}(T') g_{\beta'}^{(d)}(T)g_{\beta'}^{(d)}(T')\right]\nonumber  \\
    & = \sum_{\beta,\beta'\in\chi_d}s^{|\beta|+|\beta'|-2} \left( \sum_{T \in \chi_d} \pzero{d}{T} g_\beta^{(d)}(T) g_{\beta'}^{(d)}(T) \right)^2 \nonumber  \\
     & = \sum_{\beta \in\chi_d}(s^2)^{|\beta|-1} \ , \label{eq_E1Ld}
\end{align}
where in the last line we used the property of orthogonality of the eigenvectors $g_\beta$ stated in Eq.~(\ref{eq_orthogonality}). The first comment to make on this result is its independence on the average degree $\lambda$, which is not a priori obvious. Moreover, it is expressed in a very combinatorial way, as it involves the enumeration of trees $\beta$ in $\chi_d$ classified according to their sizes $|\beta|$. This combinatorial problem was the object of Otter's paper~\cite{otter_number_1948}, see also Appendix~\ref{app:modified-otter} for more details. Let us call $A_{d,n}$ the number of unlabeled trees of size $n$ and of depth at most $d$, and introduce the generating function $\psi_d(x) =\sum_{n\geq1} A_{d,n} x^{n-1} $, in such a way that $\E_1^{(d)}[ L^{(d)} ] =\psi_d(s^2)$. When $d$ increases for fixed $n$, the sequence $A_{d,n}$ increases monotonically towards its limit $A_n$, the total number of unlabeled trees of size $n$ (since such a tree has depth at most $n-1$ the sequence is actually constant for $d \ge n-1$); similarly for a fixed $x \ge 0 $ the sequence $\psi_d(x)$ increases towards $\psi(x)$, the generating function of the $A_n$. By definition the radius of convergence of $\psi$ is the Otter's constant $\alpha$, which shows that for $s > \sqrt{\alpha}$ and all $\lambda$ one has $\E_1^{(d)}[ L^{(d)} ] \to \infty$ as $d \to \infty$. Since $\E_0^{(d)}[ L^{(d)} ] = 1$ for all $d$, the condition $s > \sqrt{\alpha}$ is thus sufficient to have a much larger average value of the scores under $\pone{d}{}$ than under $\pzero{d}{} \otimes \pzero{d}{} $, for large $d$. From the numerical determination $\alpha\approx0.3383$ performed in~\cite{otter_number_1948} one finds the corresponding numerical value of the threshold on the correlation parameter, $\sqrt{\alpha} \approx 0.581$.

It would however be too naive to deduce from this statement that partial recovery is possible as soon as $s > \sqrt{\alpha}$ (and incidentally for low enough $\lambda$ this would violate the necessary condition $\lambda s > 1$ from~\cite{pmlr-v134-ganassali21a}): the average value of a random variable can be dominated by contributions of very rare events where it takes a much larger value than the typical ones. As a matter of fact the correct necessary and sufficient criterion for the possibility of one-sided detection on trees based on the computation of the likelihood ratio, and hence of partial recovery on graphs (via a slight variation of the score-based algorithm), was shown in~\cite{ganassali_correlation_2022} to be the divergence with $d$ of the Kullback-Leibler divergence between $\pone{d}{}$ and $\pzero{d}{} \otimes \pzero{d}{}$. Let us recall the definition of this measure of the difference between two probability distributions:
\begin{equation}
{\rm KL}(\mathbb{P}_1^{(d)}||\mathbb{P}_0^{(d)}\otimes \mathbb{P}_0^{(d)})=\sum_{T,T'}\mathbb{P}_1^{(d)}[T,T']\ln\left(\frac{\mathbb{P}_1^{(d)}[T,T']}{\mathbb{P}_0^{(d)}[T]\mathbb{P}_0^{(d)}[T']}\right)=\sum_{T,T'}\mathbb{P}_1^{(d)}[T,T']\ln L^{(d)}(T,T')=\E_1^{(d)}[\ln L^{(d)}] \ .
\label{eq:kullback-leibler}
\end{equation}
The last expression is in line with our previous discussion of rare events, since it corresponds to what is called a quenched average in the statistical mechanics terminology, to be contrasted with the annealed average $\ln \E_1^{(d)}[L^{(d)}]$ that can be dominated by rare, very large contributions. For finite values of the average degree $\lambda$ there does not seem to be a simple analytical expression for this quenched average. Jensen's inequality allows to bound it as $\E_1^{(d)}[\ln L^{(d)}]  \le \ln \E_1^{(d)}[ L^{(d)} ]= \ln \psi_d(s^2)$. For $s < \sqrt{\alpha}$ one can continue this chain of inequality with $\E_1^{(d)}[\ln L^{(d)}] \le \ln \psi(s^2) < \infty$ for all $d$, which implies that the threshold $s_{\rm algo}(\lambda)$ for the score-based algorithm to achieve partial recovery is lower bounded by $\sqrt{\alpha}$ for all finite $\lambda$. For $s > \sqrt{\alpha}$ the divergence of the annealed average with $d$ does not in general imply the growth of the quenched one; further analytical progress can however be made by considering the large $\lambda$ limit. It has indeed been shown in~\cite{ganassali_statistical_2022} that in this limit Gaussian simplifications (somehow reminiscent of the Central Limit Theorem) arise and yield
\begin{equation}
  \lim_{\lambda \to \infty} \E_1^{(d)}[\ln L^{(d)}] = \frac{1}{2} \ln \E_1^{(d)}[L^{(d)}] \ .
  \label{eq_E1L1_Gaussian}
\end{equation}
This implies the divergence of the Kullback-Leibler divergence (\ref{eq:kullback-leibler}) for $s > \sqrt{\alpha}$ if the limit $d \to \infty$ is taken after $\lambda \to \infty$. With some additional work this allowed to conclude in~\cite{ganassali_statistical_2022} that the algorithmic phase transition line $s_{\rm algo}(\lambda)$ converges to Otter's threshold $\sqrt{\alpha}$ in the large $\lambda$ limit.

Let us now turn to the finite $m$ version of the algorithm, in which the scores are computed with $S^{(d)} = \hL^{(d)}$, the function defined in Eq.~(\ref{eq_def_hLd}). Parts of the previous discussion can be adapted to this case, in particular the computation of the average of the scores under the correlated and uncorrelated laws. For the latter one has from (\ref{eq_def_hLd}):
\begin{align}
\E_0^{(d)}[ \hL^{(d)} ]& =
                         \sum_{\beta\in\hchi_d} s^{|\beta|-1} \sum_{T,T' \in \chi_d} \pzero{d}{T}\pzero{d}{T'}  g_\beta^{(d)}(T) g_\beta^{(d)}(T')  \nonumber \\
                       & =
\sum_{\beta\in\hchi_d} s^{|\beta|-1} \left( \sum_{T \in \chi_d} \pzero{d}{T}  g_\beta^{(d)}(T) g_\bullet^{(d)}(T) \right)^2 \nonumber \\
& = \sum_{\beta\in\hchi_d} s^{|\beta|-1} \delta_{\beta,\bullet} \nonumber \\
& = 1 \ ,
\end{align}
where we used $g_\bullet^{(d)}(T) =1$ for the trivial tree with a unique vertex, the orthogonality between eigenvectors stated in Eq.~(\ref{eq_orthogonality}), the fact that $|\bullet| =1$, and that $\bullet \in \hchi_d$. One thus sees that the approximate scores share with the likelihood ratio their average value under the uncorrelated law, for any value of $m$. The first moment of $\hL^{(d)}$ under $\pone{d}{}$ can be obtained by a simple adaptation of the computation done above in Eq.~(\ref{eq_E1Ld}):
\begin{align}
\E_1^{(d)}[ \hL^{(d)} ]& =\E_0^{(d)}[L^{(d)} \hL^{(d)} ] \nonumber \\
& = \sum_{\beta \in\chi_d ,\beta'\in\hchi_d} s^{|\beta|+|\beta'|-2} \, \E_0^{(d)} \left[g_\beta^{(d)}(T)g_\beta^{(d)}(T') g_{\beta'}^{(d)}(T)g_{\beta'}^{(d)}(T')\right]\nonumber \\
& = \sum_{\beta \in\chi_d ,\beta'\in\hchi_d} s^{|\beta|+|\beta'|-2} \left( \sum_{T \in \chi_d} \pzero{d}{T} g_\beta^{(d)}(T) g_{\beta'}^{(d)}(T) \right)^2 \nonumber \\
& = \sum_{\beta \in\hchi_d}(s^2)^{|\beta|-1} =\hpsi_d(s^2) \ .
\label{eq_E1hL}
\end{align}
In the last line we have introduced the generating function $\hpsi_d$ that is the counterpart of $\psi_d$ when the set $\chi_d$ of rooted unlabeled trees of depth at most $d$ is replaced by its subset $\hchi_d$, in which one furthermore imposes that all vertices have at most $m$ offsprings. In formula this means that $\hpsi_d(x) =\sum_{n\geq1} \hA_{d,n} x^{n-1} $, with $\hA_{d,n}$ the number of elements of $\hchi_d$ with $n$ vertices. Note as a side remark that a similar computation shows that $\E_0^{(d)}[(\hL^{(d)})^2] = \E_0^{(d)}[L^{(d)} \hL^{(d)} ]$, and as a consequence that $\E_0^{(d)}[(L^{(d)}-\hL^{(d)})^2]$ decreases monotonically with $m$; this interplay between the quality of the approximation of the likelihood ratio and the complexity of its computation bears similarities with the low-degree polynomial approach to hypothesis testing problems~\cite{kunisky_notes_2019}. However, we are interested in this paper to an estimation problem, namely the recovery of $\pi_\star$, for which the hypothesis testing problem on trees serves only as an intermediate step, hence as we mentioned previously this improvement of the mean square error of the approximation of $L^{(d)}$ has no direct translation into the overlap between the groundtruth $\pi_\star$ and its estimation $\hpi$.

The numbers $\hA_{d,n}$ and the associated generating functions $\hpsi_d$ have qualitative behaviors similar to those of $A_{d,n}$ and $\psi_d$: when $d$ increases $\hA_{d,n} \to \hA_n$ and for $x\ge 0$ we have $\hpsi_d(x) \to \hpsi(x)$, both limits being monotonically increasing. This modified enumeration problem was actually considered in Otter's original paper~\cite{otter_number_1948}, which established the functional equations satisfied by $\hpsi$ (see Appendix~\ref{app:modified-otter} for more details) and computed numerically their radius of convergence $\halpha$ for some small values of $m$. The corresponding thresholds $\sqrt{\halpha}$ are thus found to be $\sqrt{\halpha} \approx 0.635$ for $m=2$ and $\sqrt{\halpha} \approx 0.596$ for $m=3$ (since there are fewer trees in $\hchi_d$ than in $\chi_d$ one has $\halpha > \alpha$).

The similarity of behavior between the scores $L^{(d)}$ and their finite $m$ version $\hL^{(d)}$, at least at the level of their first moments under $\pzero{d}{} \otimes \pzero{d}{}$ and $\pone{d}{}$, brings us to the conjecture that the large $\lambda$ limit of the transition line $\widehat{s}_{\rm algo}(\lambda)$ is the Otter's threshold $\sqrt{\halpha}$ (for the corresponding value of $m$). We believe indeed that, in this large $\lambda$ limit, the same kind of simplifications that induced (\ref{eq_E1L1_Gaussian}) in the $m=\infty$ case are also at play, and implies that the divergence with $d$ of the typical values of $\hL^{(d)}$ under $\pone{d}{}$ occurs if and only if the average value computed in (\ref{eq_E1hL}) diverges. One could argue that the large $\lambda$ limits of the numerical results presented in the right panel of Fig.~\ref{fig:overlaps-2vs3_and_phase-diag-approx-vs-full} do not agree with this conjecture; recall however that they suffer from strong finite $n$ corrections, and from the arbitrariness of the $R>0$ threshold we had to use in this analysis. To make a fair comparison one should point out that the asymptotic value $\sqrt{\alpha} \approx 0.58$ for the $m=\infty$ algorithm, proven rigorously in~\cite{ganassali_statistical_2022}, is not really apparent from the corresponding curve in Fig.~\ref{fig:overlaps-2vs3_and_phase-diag-approx-vs-full}, hence we do not think that numerical simulations will bring convincing arguments in favor or against the conjecture.

\begin{figure}
\centering
\includegraphics[width=.49\textwidth]{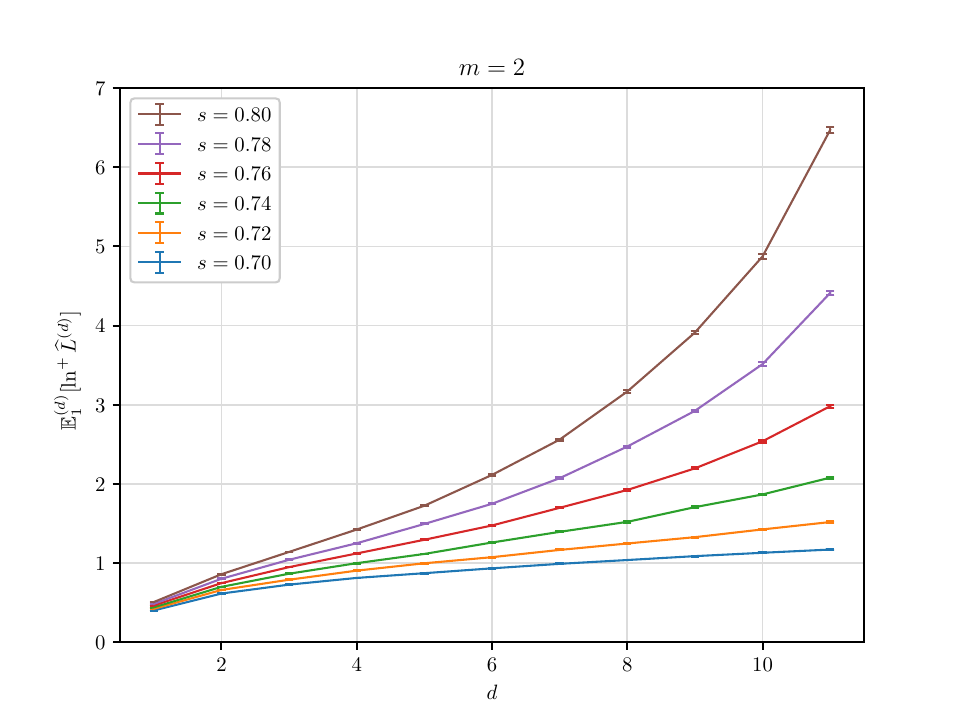}
\includegraphics[width=.49\textwidth]{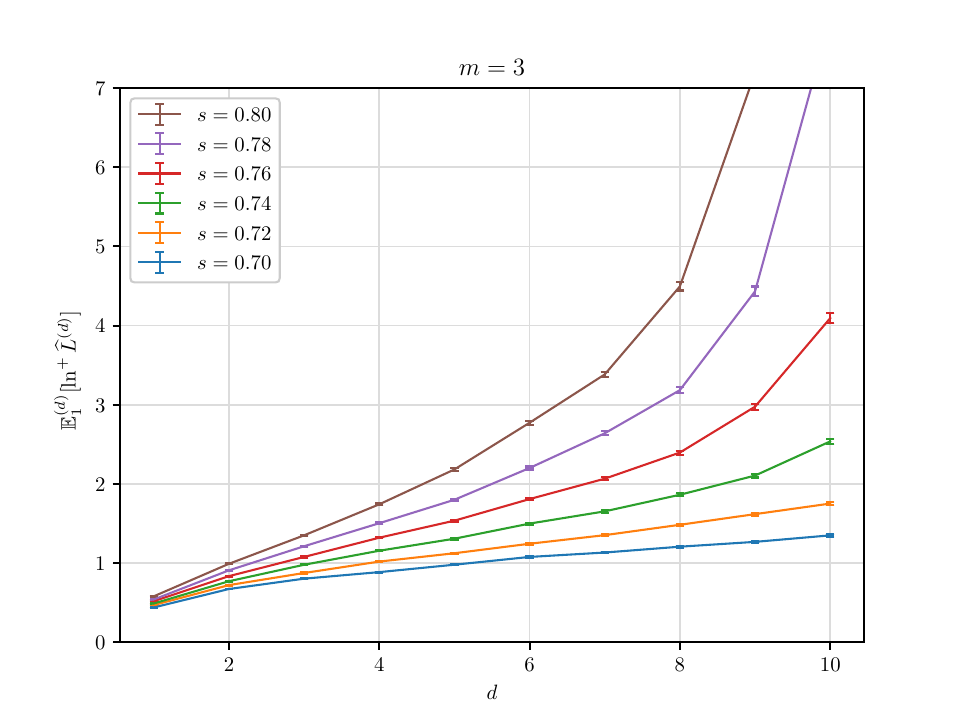}
\caption{The dependency of $\mathbb{E}_1^{(d)}[\ln^+\hL^{(d)}]$ on the depth $d$ for $m=2$ (left panel) and $m=3$ (right panel) with $\lambda=2.1$ and several values of $s$; each point is an average over $2\times10^4$ samples for $m=2$ and $5\times10^4$ samples for $m=3$. A rough estimate of the critical value $s_{\rm c}(\lambda,m)$ above which these curves diverge with $d$ as a change of curvature yields $s_{\rm c}(2.1,2) \approx 0.74$ and $s_{\rm c}(2.1,3) \approx 0.72$, while the $m=\infty$ analysis of~\cite{piccioli_aligning_2022} gave $s_{\rm c}(2.1) \approx 0.66$.
}
\label{fig:kld}
\end{figure}

A proof of the conjecture should instead rely on analytical computations; we provide in Appendices~\ref{app:martingale} and \ref{app_largedeg} some initial steps in this direction. In the former we show that a technical property of $L^{(d)}$ that was crucially used in~\cite{ganassali_correlation_2022}, namely its martingale character, is also valid for $\hL^{(d)}$, while in the latter we give an expression of the $\lambda \to \infty$ limit of $\hL^{(d)}$ where some Gaussian simplifications are performed. Unfortunately the result is still rather complicated and we did not succeed in concluding the proof. An additional difficulty with respect to the $m=\infty$ case which was handled in~\cite{ganassali_statistical_2022} is the fact that $\hL^{(d)}$ can take negative values, hence its ``typical'' value cannot be computed as $\E_1^{(d)}[\ln \hL^{(d)}]$, which is not well-defined. A natural way to bypass this problem would be to define a truncated version of the logarithm, 
\begin{equation}
\ln^+ x =
\begin{cases}
\ln x &\text{for $x>1$}\\
0 &\text{otherwise}
\end{cases} \ ,
\label{eq_def_lnplus}
\end{equation}
and to consider instead the behavior with $d$ of $\E_1^{(d)}[\ln^+ \hL^{(d)}]$ (the threshold 1 in the definition (\ref{eq_def_lnplus}) could probably be replaced by any strictly positive constant since we are interested in the tail of the distribution). We present in Fig.~\ref{fig:kld} numerical determinations of this quantity, obtained through brute force computations where we generated a large number of pairs of correlated trees $(T,T')$ and computed $\hL^{(d)}(T,T')$ by a recursion towards their roots. Their qualitative behavior is the same as the one of $\E_1^{(d)}[\ln L^{(d)}]$ that was studied numerically in~\cite{piccioli_aligning_2022}: they are increasing functions of $d$, that converges to a finite value for $s$ smaller than a threshold $s_{\rm c}(\lambda,m)$, and diverge (very fast) with $d$ when $s > s_{\rm c}(\lambda,m)$. An accurate determination of this threshold is rather difficult: the range of $d$ that is accessible to numerical simulations is limited by the exponential growth of the size of the trees with $d$, hence extrapolations to the infinite $d$ limit are somehow shaky. A stronger conjecture would be a finite $\lambda$ criterion for the algorithm to perform partial recovery, namely $\widehat{s}_{\rm algo}(\lambda) = s_{\rm c}(\lambda,m)$. We are less convinced by this conjecture at finite $\lambda$, some numerical results obtained with $\lambda$ slightly above 1 suggest indeed that $\lambda s_{\rm c}(\lambda,m) < 1$, which would violate the criterion $\lambda s > 1$ for the information-theoretic possibility of partial recovery~\cite{pmlr-v134-ganassali21a} if one had $\widehat{s}_{\rm algo}(\lambda) = s_{\rm c}(\lambda,m)$.

\section{Conclusions}
\label{sec_conclu}

We have introduced in this work variants of the message-passing algorithm of~\cite{ganassali_correlation_2022,piccioli_aligning_2022} for the partial alignment of sparse correlated ER graphs that have a lower computational cost, and demonstrated through numerical simulations that their prediction accuracy is only slightly reduced. Some analytical computations brought us to conjectures on their phase transitions in the large degree limit, through combinatorial enumerations of well-chosen families of trees.

We consider the task of proving, or disproving, these conjectures as a challenging open problem for future work. More generically, the understanding of the qualitative and quantitative connections between the graph alignment problem and its local limit, i.e. the hypothesis testing of correlation in trees, is much less satisfactory than in other inference problems on graphs, like the stochastic block model one, and would deserve to be further improved. Even for the better controlled original algorithm of~\cite{ganassali_correlation_2022,piccioli_aligning_2022} the connection has been established only at a qualitative level, with an equivalence between the partial recovery on graphs and the one-sided detection on trees, but not at a quantitative one that would yield a prediction for the overlap between the estimated permutation $\hpi$ and the ground truth $\pi_\star$. For the approximations introduced in this work a similar qualitative criterion should constitute a first objective in this research direction.

Another question that remains open and worth investigating is the status of the hard phase in the sketch of Fig.~\ref{fig_sketch_pd}: if it seems reasonable to assume that the algorithm based on the likelihood-ratio is the optimal one among those that use only local information on the graph, given the Bayesian derivation presented in Sec.~\ref{sec:original-alg}, it is not clear whether or not some global information could be exploited in polynomial time, for instance through a spectral algorithm on a carefully designed matrix, and enlarge the easy phase of this problem.

\section*{Acknowledgments}

GS warmly thanks Luca Ganassali, Marc Lelarge, Laurent Massouli\'e, Giovanni Piccioli, Gabriele Sicuro and Lenka Zdeborov\'a for fruitful previous collaborations and discussions on this topic. This work has been done within the bilateral exchange program between \'Ecole Normale Sup\'erieure - ENS, Paris (France) and Scuola di Studi Superiori ``Ferdinando Rossi'' dell'Universit\`a degli Studi di Torino - SSST, Turin (Italy).

\bibliography{biblio2}

\appendix

\section{Technical details on the computations of the approximate scores}

We provide in this Appendix some additional details on the computations presented in the main part of the text.

\subsection{The recursive definition of the eigenvectors}
\label{app_ev_definitions}

We shall first give an explicit definition of the eigenvectors $g^{(d)}_\beta$ that appeared in the main text in Eq.~(\ref{eq:theL2formula}), as a building block of the decomposition of the likelihood ratio. These objects are functions from $\chi_d$ to $\mathbb{R}$, indexed by $\chi_d$, where we recall that $\chi_d$ denotes the set of unlabeled rooted trees of depth at most $d$. They are defined by induction on $d$; for $d=0$ the only tree in $\chi_d$ is the trivial one which contains only the root, denoted $\bullet$, and one sets $g_\bullet^{(0)}(\bullet)=1$. For the induction step we shall give an expression of  $g_\gamma^{(d+1)}(N)$, where $\gamma$ and $N$ in $\chi_{d+1}$ are represented respectively as $\{\gamma_\beta\}_{\beta \in \chi_d}$ and $\{N_T\}_{T \in \chi_d}$, these vectors of non-negative integers giving the number of copies of distinct subtrees $\beta$ and $T$ rooted at the offsprings of the roots of $N$ and $\gamma$ (recall the definition given in Sec.~\ref{sec_basic_def} and illustrated in Fig.~\ref{fig_def_trees}). Let us introduce two families of formal variables, $u=\{u_\beta\}_{\beta \in \chi_d}$ and $v=\{v_T\}_{T \in \chi_d}$, and generalize the notation already used for the extraction of formal power series to $[u^\gamma] = [\prod_\beta u_\beta^{\gamma_\beta}]$ and $[v^N] = [\prod_T v_T^{N_T}]$; the products run here over $\chi_d$, without any convergence problem since for finite trees only a finite number of the $N_T$'s and $\gamma_\beta$'s are non-zero. With these conventions we can state three equivalent forms of the induction (see~\cite{ganassali_statistical_2022} for more details):
\begin{align}
  g_\gamma^{(d+1)}(N) & = \sqrt{\prod_\beta \gamma_\beta ! } [u^\gamma] \exp \left( - \sqrt{\lambda} \sum_{\beta,T} u_\beta \pzero{d}{T} g_\beta^{(d)}(T)\right) \prod_T \left(1 + \frac{1}{\sqrt{\lambda}} \sum_\beta u_\beta g_\beta^{(d)}(T)  \right)^{N_T} \label{eq:eigenvectors-alternative} \\
  & = \sqrt{\prod_\beta \gamma_\beta ! } \prod_T N_T! [u^\gamma v^N] \exp \left( - \sqrt{\lambda} \sum_{\beta,T} u_\beta \pzero{d}{T} g_\beta^{(d)}(T) + \sum_T v_T + \frac{1}{\sqrt{\lambda}} \sum_{\beta , T} u_\beta v_T g_\beta^{(d)}(T) \right) \label{eq:eigenvectors-alternative_inter} \\
  & =\frac{1}{\sqrt{\underset{\beta}{\prod} \gamma_\beta ! }} \prod_T N_T!
[v^N] \exp \left( \sum_T v_T \right) \prod_{\beta}\left(\sum_T g_{\beta}^{(d)}(T)\left(\frac{v_T}{\sqrt{\lambda}}-\sqrt{\lambda}\pzero{d}{T} \right)\right)^{\gamma_\beta} \ .
	\label{eq:eigenvectors-alternative2}
\end{align}
In these expressions all the sums and products on $\beta$ and $T$ run over $\chi_d$; as already mentioned there are no problems of convergence since only a finite number of terms are non-trivial in each of them. Note that in the form (\ref{eq:eigenvectors-alternative}) the dependency on $N$ is rather simple since it factorizes as a product over $T$ of terms depending exponentially on $N_T$, while the dependency on $\gamma$ lies in the coefficient extraction term $[u^\gamma]$ (apart from the normalization coefficient in front of it). The roles of $N$ and $\gamma$ are somehow reversed in the form (\ref{eq:eigenvectors-alternative2}), while the second line (\ref{eq:eigenvectors-alternative_inter}) provides a simple bridge between these two equivalent forms.

It is instructive to work out the explicit form of $g^{(d+1)}_\gamma$ for some simple dual trees $\gamma$. Consider first the simplest case $\gamma=\bullet$, the trivial tree made of only the root, which corresponds to $\gamma_\beta=0$ for all $\beta \in \chi_d$. Computing the constant term in the formal power series of (\ref{eq:eigenvectors-alternative}) by setting $u=0$ one finds immediately $g_\bullet^{(d+1)}(N)=1$. As a second example suppose that in $\gamma$ the root has $q>0$ offsprings, the latter being without any descendants, that is $\gamma_\beta = q \one(\beta=\bullet)$. Hence, denoting more simply $u=u_\bullet$ the only formal variable that plays a non-trivial role, one obtains from (\ref{eq:eigenvectors-alternative}) for this $\gamma$:
\begin{align}
  g_\gamma^{(d+1)}(N) & = \sqrt{q!} [u^q] \exp \left( - u \sqrt{\lambda} \sum_{T}  \pzero{d}{T} g_\bullet^{(d)}(T)\right) \prod_T \left(1 + \frac{1}{\sqrt{\lambda}}  u g_\bullet^{(d)}(T)  \right)^{N_T} \\
& = \sqrt{q!} [u^q] e^{-u \sqrt{\lambda}} \left(1 + \frac{u}{\sqrt{\lambda}} \right)^{\sum_T N_T} \\
& = \mathcal{C}_q(l) \ ,
\end{align}
where we used the fact $g_\bullet^{(d)}(T)=1$ previously established, denoted $l = \sum_T N_T$ the degree of the root of $N$, and defined the $q$-th Charlier polynomial 
\beq
\mathcal{C}_q(x) = \sqrt{q!} [u^q] e^{-u \sqrt{\lambda}} \left(1 + \frac{u}{\sqrt{\lambda}} \right)^x \ .
\label{eq_def_Charlier}
\eeq
These form the family of orthogonal polynomials for the Poisson distribution, in the sense that $\mathcal{C}_q(x)$ is a polynomial in $x$ of degree $q$, with the orthogonality property:
\beq
\sum_{l=0}^\infty e^{-\lambda} \frac{\lambda^l}{l!} \mathcal{C}_q(l) \mathcal{C}_{q'}(l) = \delta_{q,q'} \ . 
\eeq
The first Charlier polynomials are easily found to be
\beq
\mathcal{C}_0(x) = 1 \ , \qquad \mathcal{C}_1(x) = \frac{x-\lambda}{\sqrt{\lambda}} \ , \qquad 
\mathcal{C}_2(x) = \frac{1}{\sqrt{2}} \frac{x(x-1) - 2 \lambda x +\lambda^2}{\lambda} \ .
\label{eq_examples_Charlier}
\eeq
Additional examples of $g_\beta$ for a few more dual trees will be presented in Appendix~\ref{app:martingale}, along with other generic properties of these eigenvectors.

\subsection{The proof of the recursive formula for $\hL$}
\label{app:proof-truncation}

We shall now prove that the function $\hL^{(d)}$ defined in Eq.~(\ref{eq_def_hLd}), with $\hchi_d$ the set of rooted unlabeled trees of depth at most $d$ in which each vertex has at most $m$ offsprings, obeys the recursive decomposition stated in Eqs.~(\ref{eq_hLd_recursion},\ref{eq:alg-func-trunc}). 

To achieve this goal let us first consider an arbitrary family $\tchi_d \subset \chi_d$ of trees of depth at most $d$, and define the following function from $\chi_d \times \chi_d$ to $\mathbb{R}$:
\beq
\tL^{(d)}(T,T')=\sum_{\beta\in \tchi_d} s^{|\beta|-1} g_\beta^{(d)}(T) g_\beta^{(d)}(T') \ .
\label{eq_tld}
\eeq
We fix now an integer $p$, and define $\tchi_{d+1} \subset \chi_{d+1}$ as the set of trees in which the root has exactly $p$ offsprings, each of them being the root of a tree in $\tchi_d$, and set for $N,N' \in \chi_{d+1}$:
\beq
\tL^{(d+1)}(N,N')=\sum_{\gamma\in \tchi_{d+1}} s^{|\gamma|-1} g_\gamma^{(d+1)}(N) g_\gamma^{(d+1)}(N') \ .
\label{eq_tldp1}
\eeq 
We claim that, if the root of $N$ (resp. $N'$) has $l$ (resp. $l'$) offsprings, themselves roots of $T_1,\dots,T_l \in \chi_d$ (resp. $T'_1,\dots,T'_{l'}$), then
\begin{equation}
  \tL^{(d+1)}(N,N')=
    s^p[t^p]e^{\lambda t}(1-t)^{l+l'}\sum_{k=0}^{\min(l,l')}\left(\frac{t}{\lambda(1-t)^2}\right)^k \sum_{I,I',\sigma}\prod_{i\in I} \tL^{(d)}(T_i,T'_{\sigma(i)}) \ ,
\label{eq:truncated-L}
\end{equation}
where $I$ (resp. $I'$) is a subset of $[l]$ (resp. $[l']$) containing $k$ elements, and $\sigma$ a bijection from $I$ to $I'$. Note that the validity of (\ref{eq_hLd_recursion},\ref{eq:alg-func-trunc}) is then a direct consequence of (\ref{eq:truncated-L}), by taking $\tchi_d=\hchi_d$ and viewing $\hchi_{d+1}$ as the disjoint union of the $\tchi_{d+1}$ when $p$ varies from $0$ to $m$; moreover by sending $m$ to infinity one also proves in this way the original recursion (\ref{eq:theLformula},\ref{eq:alg-func}).

We start now the proof of (\ref{eq:truncated-L}). To deal with the definition in (\ref{eq_tldp1}) we have to perform a sum over $\gamma \in \tchi_{d+1}$; the tree $\gamma$ is represented as $\{\gamma_\beta\}_{\beta \in \chi_d}$, with $\gamma_\beta$ the number of copies of $\beta$ rooted at the offsprings of the root of $\gamma$. By construction these subtrees are in $\tchi_d$, hence $\gamma_\beta=0$ if $\beta \in \chi_d \setminus \tchi_d$; moreover the degree of the root of $\gamma$ is fixed to $p$ by the definition of $\tchi_{d+1}$. Noting that the sizes of the trees obey the identity $|\gamma| = 1+ \sum_{\beta} \gamma_\beta |\beta|$, we can write for
an arbitrary function $h$ defined on $\tchi_{d+1}$:
\begin{align}
  \sum_{\gamma \in \tchi_{d+1}} s^{|\gamma|-1} h(\gamma) & = \sum_{\{\gamma_\beta \ge 0\}_{\beta\in\tchi_d}} s^{\sum_\beta \gamma_\beta |\beta|} \, \one ( {\textstyle\sum}_\beta \gamma_\beta = p) \, h(\gamma) \label{eq_summation_tchi}\\
& = \sum_{\{\gamma_\beta \ge 0\}_{\beta\in\tchi_d}} s^{\sum_\beta \gamma_\beta} s^{\sum_\beta \gamma_\beta (|\beta|-1)} \, \one({\textstyle\sum}_\beta \gamma_\beta = p) \,  h(\gamma) \nonumber \\
& = s^p [t^p] \sum_{\{\gamma_\beta \ge 0\}_{\beta\in\tchi_d}} \prod_{\beta\in\tchi_d}\left(t s^{|\beta| -1}\right)^{\gamma_\beta} h(\gamma) \ . \nonumber
\end{align}
Let us now apply this formula with $h(\gamma)=g_\gamma^{(d+1)}(N)g_\gamma^{(d+1)}(N')$, which corresponds to the function appearing in (\ref{eq_tldp1}). Using the expression stated in Eq.~(\ref{eq:eigenvectors-alternative2}) for $g_\gamma^{(d+1)}$, one gets
\begin{align}
\tL^{(d+1)}(N,N')  = &  s^p[t^p] \left(\prod_{T\in\chi_d} N_T!N'_T!\right) [v^Nv'^{N'}] \exp\left(\sum_{T\in \chi_d} (v_T+v'_T) \right) \nonumber \\
                  & \prod_{\beta\in\tchi_d} \left\{\sum_{\gamma_\beta=0}^\infty \frac{1}{\gamma_\beta!}\left[t s^{|\beta|-1}\sum_{T,T' \in \chi_d} g_\beta^{(d)}(T)g_\beta^{(d)}(T')\left(\frac{v_T}{\sqrt\lambda}-\sqrt\lambda\pzero{d}{T}\right)\left(\frac{v'_{T'}}{\sqrt\lambda}-\sqrt\lambda \pzero{d}{T'} \right)\right]^{\gamma_\beta}\right\} \nonumber \\
   = &  s^p[t^p] \left(\prod_{T\in\chi_d} N_T!N'_T!\right) [v^Nv'^{N'}] \exp\left(\sum_{T\in \chi_d} (v_T+v'_T) \right) \nonumber \\
                  & \exp \left(\sum_{\beta\in\tchi_d} t s^{|\beta|-1} \left[
                    \frac{1}{\lambda} \sum_{T,T' \in \chi_d} g_\beta^{(d)}(T)g_\beta^{(d)}(T') v_T v'_{T'} - \left(\sum_{T \in \chi_d}g_\beta^{(d)}(T) (v_T+v'_T)  \right) \left(\sum_{T \in \chi_d}\pzero{d}{T} g_\beta^{(d)}(T) \right) \right. \right. \nonumber \\
  & \hspace{2cm} \left. \left. + \lambda \left(\sum_{T \in \chi_d}\pzero{d}{T} g_\beta^{(d)}(T) \right)^2
                    \right] \right) \ . \nonumber
\end{align}
The orthogonality of the eigenvectors stated in Eq.~(\ref{eq_orthogonality}), combined with their value when $\beta=\bullet$ the trivial tree, namely $g_\bullet^{(d)}(T)=1$, implies that $\sum_{T \in \chi_d}\pzero{d}{T} g_\beta^{(d)}(T) = \one(\beta = \bullet)$. Noting that $|\bullet|=1$ we can thus simplify the last equality in
\beq
\tL^{(d+1)}(N,N') =  s^p[t^p] \left(\prod_{T\in\chi_d} N_T!N'_T!\right) [v^N v'^{N'}] \exp\left( \lambda t + (1-t) \sum_{T\in \chi_d} (v_T+v'_T) + \frac{t}{\lambda} \sum_{T,T' \in \chi_d} v_T v'_{T'} \tL^{(d)}(T,T') \right) \ ; \nonumber
\eeq
we recall that $\tL^{(d)}$ was defined in Eq.~(\ref{eq_tld}). In order to proceed we remark that the extraction of coefficients in a formal power series can be written as a partial derivative, with an appropriate combinatorial factor. For a single variable this reads $p! [x^p] h(x) = \left. \left(\frac{\partial}{\partial x} \right)^p h(x) \right|_{x=0}$, the multi-dimensional generalization that we shall use being
\beq
\left(\prod_{T\in\chi_d} N_T!N'_T!\right) [v^N v'^{N'}] h(v,v') = \left. \left( \prod_{T\in\chi_d} \left(\frac{\partial}{\partial v_T} \right)^{N_T} \left(\frac{\partial}{\partial v'_T} \right)^{N'_T} \right) h(v,v') \right|_{v=v'=0} \ .
\eeq
Since $N$ and $N'$ are described as lists $T_1,\dots,T_l$ and $T'_1,\dots,T'_{l'}$ of subtrees, one has
\beq
N_T = \sum_{i=1}^l \one(T=T_i) \ , \qquad
N'_T = \sum_{i=1}^{l'} \one(T=T'_i) \ .
\eeq
Combining these observations yields:
\begin{align}
  \tL^{(d+1)}&(N,N') =  s^p[t^p] e^{\lambda t} \left. \left( \prod_{i=1}^l \frac{\partial}{\partial v_{T_i}} \right) \left(\prod_{i'=1}^{l'} \frac{\partial}{\partial v'_{T'_{i'}}}\right)  \exp\left( (1-t) \sum_{T\in \chi_d} (v_T+v'_T) + \frac{t}{\lambda} \sum_{T,T' \in \chi_d} v_T v'_{T'} \tL^{(d)}(T,T') \right) \right|_{v=v'=0} \nonumber \\
       &=  s^p[t^p] e^{\lambda t} \left. \left( \prod_{i=1}^l \frac{\partial}{\partial v_{T_i}} \right)  e^{(1-t) \sum_T v_T}  \left( \left. \left( \prod_{i'=1}^{l'} \frac{\partial}{\partial v'_{T'_{i'}}} \right)  e^{ (1-t) \sum_T v'_T + \frac{t}{\lambda} \sum_{T,T'} v_T v'_{T'} \tL^{(d)}(T,T')} \right|_{v'=0} \right) \right|_{v=0} \ .   \label{eq_tLd_proof}
\end{align}
One can perform in a first time the derivatives with respect to $v'$ to obtain
\begin{align}
  & \left. \left( \prod_{i'=1}^{l'} \frac{\partial}{\partial v'_{T'_{i'}}} \right)   e^{ (1-t) \sum_T v'_T + \frac{t}{\lambda} \sum_{T,T'} v_T v'_{T'} \tL^{(d)}(T,T')} \right|_{v'=0} \nonumber \\ & \hspace{1cm} = \prod_{i'=1}^{l'} \left( (1-t) + \frac{t}{\lambda} \sum_{T \in \chi_d} v_T \tL^{(d)}(T,T'_{i'}) \right)
\nonumber \\
  & \hspace{1cm} = (1-t)^{l'} \sum_{I' \subset [l']} \left(\frac{t}{\lambda(1-t)}\right)^{|I'|} \sum_{\{U_{i'} \in \chi_d\}_{i'\in I'}} \prod_{i' \in I'} v_{U_{i'}} \tL^{(d)}(U_{i'},T'_{i'}) \ , \label{eq_deriv_vp}
\end{align}
where in the last line we expanded the product by introducing a sum over all subsets $I'$ of $[l']$. It remains now to observe that the derivatives with respect to $v$ are of the form:
\begin{align}
\left. \left( \prod_{i=1}^l \frac{\partial}{\partial v_{T_i}} \right) e^{(1-t) \sum_T v_T} \prod_{i' \in I'} v_{U_{i'}} \right|_{v=0} & = \sum_{I \subset [l]} (1-t)^{l-|I|}  \left. \left( \prod_{i \in I}  \frac{\partial}{\partial v_{T_i}} \right) \prod_{i' \in I'} v_{U_{i'}} \right|_{v=0} \nonumber \\
& = \sum_{I \subset [l]} (1-t)^{l-|I|}  \one(|I| = |I'|) \sum_{\sigma} \prod_{i \in I} \one(T_i=U_{\sigma(i)}) \ ; \label{eq_deriv_v}
\end{align}
in the first line we introduced a summation over $I \subset [l]$, the derivatives with respect to $T_i$ with $i \in I$ acting on the product over $i'$, the others on the exponential. The non-vanishing terms in the right-hand side of the first line are those in which each derivative with respect to $T_i$ is paired with a term $v_{U_{i'}}$ of the product, which imposes the equality of the cardinalities of $I$ and $I'$, and the existence of a bijection $\sigma$ from $I$ to $I'$, introduced in the second line. Inserting (\ref{eq_deriv_vp}) and (\ref{eq_deriv_v}) in (\ref{eq_tLd_proof}), and denoting $k=|I|=|I'|$ the common cardinality of the selected subsets of $[l]$ and $[l']$, concludes the proof of (\ref{eq:truncated-L}).

\subsection{Explicit formulas for the computation of $\hL$}
\label{app:general-approximation}

We justify here the formula (\ref{eq:alg-func-m2}) for the computation of $\hL$ in the case $m=2$, present its counterpart for $m=3$, and give some elements of a generalization to arbitrary values of $m$. Let us start by rewriting the recursion of Eq.~(\ref{eq:alg-func-trunc}) as
\beq
\hf(l,l';\{\hL_{i,i'}\})=\sum_{p=0}^m s^p[t^p]e^{\lambda t}(1-t)^{l+l'}\sum_{k \geq 0}\left(\frac{t}{\lambda(1-t)^2}\right)^k S_k \ ,
\eeq
where we defined 
\beq
S_k=S_k(l,l';\{\hL_{i,i'}\}) = \sum_{I,I',\sigma}\prod_{i\in I}\hL_{i,\sigma(i)} \ ,
\label{eq_def_Sk}
\eeq
with $I$ (resp. $I'$) a $k$-element subset of $[l]$ (resp. $[l']$), and $\sigma$ a bijection from $I$ to $I'$. By convention $S_0=1$, and $S_k(l,l';\{\hL_{i,i'}\}) = 0$ if $k > \min(l,l')$. One can further write
\beq
\hf(l,l';\{\hL_{i,i'}\})=\sum_{p=0}^m s^p \sum_{k=0}^p D_{p,k} S_k \ ,
\eeq
where the coefficients $D_{p,k} = D_{p,k}(\lambda,l,l')$ can be written explicitly, for $p \geq k$, as
\begin{align}
D_{p,k} & = [t^p]e^{\lambda t}(1-t)^{l+l'} \left(\frac{t}{\lambda(1-t)^2}\right)^k  \nonumber \\
& = \frac{1}{\lambda^k} [t^{p-k}]e^{\lambda t}(1-t)^{l+l'-2k} \nonumber \\
& =  \frac{1}{\lambda^k} \sum_{r=0}^{p-k} \frac{\lambda^r}{r!} (-1)^{p-k-r} \binom{l+l'-2k}{p-k-r} \ .
\label{eq_Dpk}
\end{align}
In particular for $m=3$ one finds
\begin{align}
\hf(l,l';\{\hL_{i,i'}\}) = 
1 & +s\left[\frac{1}{\lambda}S_1+\lambda-l-l'\right] \label{eq_hf_appendix} \\ &+\frac{s^2}{2}\left[\lambda^2-2\lambda(l+l')+(l+l')(l+l'-1)+2\left(1-\frac{l+l'-2}{\lambda}\right)S_1+\frac{2}{\lambda^2}S_2\right]  \nonumber \\
	&+\frac{s^3}{6}\bigg[\lambda^3-3\lambda^2(l+l')+3\lambda(l+l')(l+l'-1)- (l+l')(l+l'-1)(l+l'-2) \nonumber \\
	&\hspace{3em}+\frac{3}{\lambda}\left(\lambda^2-2\lambda(l+l'-2)+(l+l'-2)(l+l'-3)\right)S_1 
+\frac{6}{\lambda^2}\left(\lambda-l-l'+4\right)S_2 + \frac{6}{\lambda^3}S_3\bigg] \ ,  \nonumber 
\end{align}
the $m=2$ case corresponding to the two first lines of this equation, and the generalization to higher values of $m$ of this formula can be obtained straightforwardly from (\ref{eq_Dpk}).

The difficulty that remains to be discussed is the efficient computation of the $S_k$'s: in principle, for a fixed $k$, the computational cost for the determination of $S_k$ (as defined in Eq.~(\ref{eq_def_Sk})) scales with $l$ and $l'$ as $O((l l')^k)$, because the number of $k$-element subsets of $[l]$ is $\binom{l}{k} = O(l^k)$, and an analog formula holds for the subsets of $[l']$. As we shall see now, the use of the inclusion-exclusion principle allows to reduce this cost for the small values of $k$.

The term $S_1$ can be immediately written down from its definition:
\beq
S_1= \sum_{i,i'}\hL_{i,i'} \ ,
\eeq
where here and in the following unspecified summations over an unprimed (resp. primed) index are understood to run over $[l]$ (resp. $[l']$). 

The term $S_2$ can be manipulated as follows:
\begin{align}
S_2& = \sum_{i_1 < i_2} \sum_{i'_1 < i'_2} (\hL_{i_1,i'_1} \hL_{i_2,i'_2} + \hL_{i_1,i'_2} \hL_{i_2,i'_1}   ) \label{eq_S2_1} \\
& =\frac{1}{2}  \sum_{i_1 \neq i_2} \sum_{i'_1 \neq i'_2} \hL_{i_1,i'_1} \hL_{i_2,i'_2} \label{eq_S2_2} \\
&=\frac{1}{2}\sum_{i_1 , i_2 , i'_1 , i'_2}  \left(1-\delta_{i_1,i_2}-\delta_{i'_1,i'_2}+\delta_{i_1,i_2} \delta_{i'_1,i'_2}\right) \hL_{i_1,i'_1} \hL_{i_2,i'_2} \\
&=\frac{1}{2}\left(\left(\sum_{i,i'}\hL_{i,i'}\right)^2-\sum_i\left(\sum_{i'}\hL_{i,i'}\right)^2-\sum_{i'}\left(\sum_i\hL_{i,i'}\right)^2+\sum_{i,i'}\left(\hL_{i,i'}\right)^2\right) \ . \label{eq_S2_4}
\end{align}
From the expression of the last line one can compute $S_2$ in $O(ll')$ operations, hence notably faster than the naive estimate $O((ll')^2)$ obtained by counting the number of terms in the first line. Inserting these expressions of $S_1$ and $S_2$ in the first two lines of (\ref{eq_hf_appendix}) completes the justification of Eq.~(\ref{eq:alg-func-m2}).

Similar manipulations can be performed for arbitrary $k$; the generalization of the step leading from (\ref{eq_S2_1}) to (\ref{eq_S2_2}) reads:
\beq
S_k  = \sum_{i_1<\dots<i_k} \sum_{i'_1<\dots<i'_k} \ \sum_{\sigma \in {\mathcal S}_k} \ \prod_{a=1}^k \hL_{i_a,i'_{\sigma(a)}} = \frac{1}{k!} \sum_{i_1,\dots , i_k}^\star  \sum_{i'_1, \dots , i'_k}^\star \prod_{a=1}^k \hL_{i_a,i'_a} \ ,
\eeq
where ${\mathcal S}_k$ denotes the group of permutations of $[k]$, and where the star on the summations means that the indices are pairwise distinct. In order to generalize the transformation that led from (\ref{eq_S2_2}) to (\ref{eq_S2_4}), i.e. the rewriting of these constrained summations in terms of unconstrained ones with repeated indices, one can use the M\"obius inversion formula in the lattice of partitions (see e.g. chapter 3 in~\cite{Stanley_vol1}). We will content ourselves with a more pedestrian derivation for $k=3$: for an arbitrary function $h$ of 3 indices (not necessarily invariant under their permutations), one has
\begin{align}
\sum_{i_1,i_2,i_3}^\star  h(i_1,i_2,i_3) &= \sum_{i_1,i_2,i_3} (1-\delta_{1,2})(1-\delta_{1,3})(1-\delta_{2,3})h(i_1,i_2,i_3)=\\
& =\underbrace{\sum_{i_1,i_2,i_3}h(i_1,i_2,i_3)}_{{(i)}}
\underbrace{-\sum_{i_1,i_2}\left(h(i_1,i_1,i_2)+h(i_1,i_2,i_1)+h(i_2,i_1,i_1)\right)}_{{(ii)}}
\underbrace{+2\sum_{i_1}h(i_1,i_1,i_1)}_{{(iii)}} \ .
\label{eq:f-delta}
\end{align}
We shall thus write formally
\beq
S_3 = \frac{1}{6} \big( (i) + (ii) + (iii) \big) \big( (i') + (ii') + (iii') \big) \ ,
\eeq
where the three terms in each factor correspond to the patterns of equality between indices as defined in Eq.~(\ref{eq:f-delta}), and the two factors correspond to the summations over $I$ and $I'$. The ``diagonal'' terms in this product are
{\allowdisplaybreaks\begin{align*}
{(i)(i')} ={}& \sum_{\substack{i_1,i_2,i_3\\i'_1,i'_2,i'_3}}\hL_{i_1,i'_1}\hL_{i_2,i'_2}\hL_{i_3,i'_3} = \left(\sum_{i,i'}\hL_{i , i'}\right)^3 \ , \\
{(ii)(ii')} ={}& \sum_{\substack{i_1,i_2\\i'_1,i'_2}} \left(3\hL_{i_1,i'_1}^2\hL_{i_1,i'_2} + 6\hL_{i_1,i'_1}\hL_{i_1,i'_2}\hL_{i_2,i'_1}\right)\\
={}&3\left(\sum_{i,i'}\hL_{i , i'}\right)\left(\sum_{i,i'}\left(\hL_{i , i'}\right)^2\right) + 6\sum_{i,i'}\left[\hL_{i , i'}\left(\sum_j\hL_{j , i'}\right)\left(\sum_{j'}\hL_{i , j'}\right)\right] \ , \\
{(iii)(iii')} ={}& 4\sum_{i,i'}\left(\hat L_{i , i'}\right)^3 \ ,
\tag{\stepcounter{equation}\theequation}
\end{align*}}
while the ``off-diagonal'' ones read
{\allowdisplaybreaks\begin{align*}
{(i)(ii')} ={}& -\sum_{\substack{i_1,i_2,i_3\\i'_1,i'_2}}\left(\hL_{i_1,i'_1}\hL_{i_2,i'_1}\hL_{i_3,i'_2} + \hL_{i_1,i'_1}\hL_{i_2,i'_2}\hL_{i_3,i'_1} + \hL_{i_1,i'_2}\hL_{i_2,i'_1}\hL_{i_3,i'_1}\right)\\
={}&-3\left(\sum_{i,i'}\hL_{i , i'}\right)\left(\sum_{i'}\left(\sum_i \hL_{i , i'}\right)^2\right) \ , \\
{(i)(iii')} ={}& 2\sum_{\substack{i_1,i_2,i_3\\i'_1}}\hL_{i_1,i'_1}\hL_{i_2,i'_1}\hL_{i_3,i'_1} = 2\sum_{i'}\left(\sum_i\hL_{i , i'}\right)^3 \ , \\
{(ii)(iii')} ={}& -6\sum_{\substack{i_1,i_2\\i'_1}}\hL_{i_1,i'_1}\hL_{i_1,i'_1}\hL_{i_2,i'_1} = -6\sum_{i'}\left(\left(\sum_i\hL_{i , i'}\right)\left(\sum_i\left(\hL_{i , i'}\right)^2\right)\right) \ ,
\tag{\stepcounter{equation}\theequation}
\end{align*}}\\
as well as the three terms obtained by exchanging primed and unprimed indices. Collecting all these contributions yields finally
\begin{equation}
\begin{split}
S_3 = \frac{1}{6}\left\{\rule{0em}{2.5em}\right.
&\left(\sum_{i,i'}\hL_{i,i'}\right)^3
-3\left(\sum_{i,i'}\hL_{i,i'}\right)\left(\sum_{i'}\left(\sum_i \hL_{i,i'}\right)^2\right)
-3\left(\sum_{i,i'}\hL_{i,i'}\right)\left(\sum_{i}\left(\sum_{i'} \hL_{i,i'}\right)^2\right)  \\
& +2\sum_{i'}\left(\sum_i\hL_{i,i'}\right)^3 + 2\sum_{i}\left(\sum_{i'}\hL_{i,i'}\right)^3
+3\left(\sum_{i,i'}\hL_{i,i'}\right)\left(\sum_{i,i'}\left(\hL_{i,i'}\right)^2\right)  \\
& +6\sum_{i,i'}\left[\hL_{i,i'}\left(\sum_j\hL_{j , i'}\right)\left(\sum_{j'}\hL_{i , j'}\right)\right]
-6\sum_{i'}\left(\left(\sum_i\hL_{i,i'}\right)\left(\sum_i\left(\hL_{i,i'}\right)^2\right)\right) \\
& -6\sum_{i}\left(\left(\sum_{i'}\hL_{i,i'}\right)\left(\sum_{i'}\left(\hL_{i,i'}\right)^2\right)\right) + 
4\sum_{i,i'}\left(\hL_{i,i'}\right)^3
\left.\rule{0em}{2.5em}\right\} \ .
\end{split}
\label{eq_S3}
\end{equation}
Along with (\ref{eq_hf_appendix}) this expression completes the presentation of the formula that we used for the numerical implementation of the $m=3$ algorithm. Note that (\ref{eq_S3}) reveals that $S_3$ can be computed in $O(ll')$ operations, as $S_2$, hence more efficiently than the $O((ll')^3)$ number of terms of its initial definition. As a matter of fact the terms of the form $\sum_{i,i'}(\hL_{i,i'})^r$, for $r \in \{1,2,3\}$ can be computed in $O(ll')$ operations, while with the bookkeeping of partial sums of the form $\sum_j (\hL_{j,i'})^r$ in two vectors indexed by $i'$ for $r \in \{1,2\}$ (and similarly for $\sum_{j'} (\hL_{i,j'})^r$), all the other terms in $S_3$ can be computed with $O(ll')$ operations.

Higher values of $k$ can in principle be treated similarly, but yield rather long and cumbersome expressions. Moreover, the scaling of the number of operations required to compute $S_k$ becomes larger than $O(ll')$ for $k>3$: for instance the expression of $S_4$ contains the term $\sum_{i',j'}\left(\sum_i\hL_{i,i'}\hL_{i,j'}\right)^2$ (coming from $\sum_{\{i_k,i'_k\}_{k\in[4]}}\delta_{i_1,i_2}\delta_{i_3,i_4}\delta_{i'_1,i'_4}\delta_{i'_2,i'_3}\hL_{i_1,i'_1}\dots\hL_{i_4,i'_4}$) which has a cost $O(l l'^2)$. This increase in complexity could be anticipated: $S_k$ becomes the permanent of a square matrix when $l=l'=k$, a quantity that is notoriously hard to compute~\cite{valiant_complexity_1979}.

\subsection{Additional properties}
\label{app:martingale}

We present in this appendix some additional properties of the eigenvectors $g_\beta^{(d)}$ and their consequences on the approximations $\hL^{(d)}$ of the likelihood ratio, in particular its martingale character.

Recall that we gave in Appendix~\ref{app_ev_definitions} the values of $g_{\gamma}^{(d+1)}(N)$ for the trivial tree $\gamma=\bullet$ and for the trees $\gamma$ of depth 1, in which the root has $q>0$ offsprings, themselves being without descendants, that we shall denote $\mathfrak{t}_q$. We found $g_{\mathfrak{t}_q}^{(d+1)}(N)=\mathcal{C}_q(l)$, with $\mathcal{C}_q$ the $q$-th Charlier polynomial defined in Eq.~(\ref{eq_def_Charlier}), and $l$ denoting the degree of the root of $N$. Using a graphical representation of the tree $\gamma$ we can write for the smallest values of $q$:
\bgroup
\def\x{0.6}
\tikzset{mynode/.append style={scale=\x}}
\tikzset{every picture/.append style={scale=\x, every node/.style=mynode, edge from parent/.style={myedge}, level 1/.style={sibling distance=10mm}, level 2/.style={sibling distance=5mm}, level distance=5mm}}
\beq
g^{(d+1)}_{\tikz{\node{};}}(N) = 1\ , \qquad
g^{(d+1)}_{\tikz{\node{} child{node{}};}}(N) = \mathcal{C}_1(l) \ , \qquad
g^{(d+1)}_{\tikz{\node{} child{node{}} child{node{}};}}(N) = \mathcal{C}_2(l) \ , 
\eeq
\egroup
the explicit values of these Charlier polynomials being given in Eq.~(\ref{eq_examples_Charlier}).

It is instructive for the following to work out some additional examples in which the tree $\gamma$ has depth 2. Suppose first that the root of $\gamma$ has exactly one offspring, which is the root of one copy of $\mathfrak{t}_q$ for some $q>0$, in such a way that $\gamma_\beta = \one (\beta =\mathfrak{t}_q)$. Denoting $u=u_{\mathfrak{t}_q}$ the only formal variable that plays a non-trivial role one obtains from (\ref{eq:eigenvectors-alternative}) for this $\gamma$: 
\begin{align}
  g_\gamma^{(d+1)}(N) & = [u] \exp \left( - u \sqrt{\lambda} \sum_{T}  \pzero{d}{T} g_{\mathfrak{t}_q}^{(d)}(T)\right) \prod_T \left(1 + \frac{1}{\sqrt{\lambda}}  u g_{\mathfrak{t}_q}^{(d)}(T)  \right)^{N_T} \\
& = [u] \prod_T \left(1 + \frac{1}{\sqrt{\lambda}}  u g_{\mathfrak{t}_q}^{(d)}(T)  \right)^{N_T} \\ 
& =  \frac{1}{\sqrt{\lambda}} \sum_T N_T g_{\mathfrak{t}_q}^{(d)}(T) \ ,
\end{align}
where in the first step we used the orthogonality between the constant function $g_\bullet^{(d)}$ and $g_{\mathfrak{t}_q}^{(d)}$ for $q >0$. Suppose that $N$ is described by the degree $l$ of its root, and by the list $T_1,\dots,T_l$ of its subtrees. We will denote $\ell_i$ the degree of the root of $T_i$; noting that $N_T = \sum_{i=1}^l \one(T=T_i)$, and using the value of $g_{\mathfrak{t}_q}^{(d)}(T)$ established in Appendix~\ref{app_ev_definitions}, one concludes that for this tree $\gamma$,
\beq
g_\gamma^{(d+1)}(N) = \frac{1}{\sqrt{\lambda}} \sum_{i=1}^l \mathcal{C}_q(\ell_i) \ .
\eeq
The graphical representations of this formula for $q \in \{1,2\}$ are
\bgroup
\def\x{0.6}
\tikzset{mynode/.append style={scale=\x}}
\tikzset{every picture/.append style={scale=\x, every node/.style=mynode, edge from parent/.style={myedge}, level 1/.style={sibling distance=10mm}, level 2/.style={sibling distance=5mm}, level distance=5mm}}
\beq
g^{(d+1)}_{\tikz{\node{} child{node{} child{node{}}};}}(N) = \frac{1}{\sqrt\lambda}\sum_{i=1}^l\mathcal{C}_1(\ell_i)\ ,
\qquad
g^{(d+1)}_{\tikz{\node{} child{node{} child{node{}} child{node{}}};}}(N) = \frac{1}{\sqrt\lambda}\sum_{i=1}^l \mathcal{C}_2(\ell_i) \ .
\eeq
\egroup
One can consider a few other examples of trees $\gamma$ of depth 2; since the computations are simple variations of the one presented above we will state directly the results. If the root of $\gamma$ has degree 2, and its two offsprings are roots of two copies of $\mathfrak{t}_q$ with $q>0$, one finds
\beq
g_\gamma^{(d+1)}(N) = \frac{1}{\lambda \sqrt{2} } \sum_{\substack{i,j=1 \\ i \neq j}}^l \mathcal{C}_q(\ell_i) \mathcal{C}_q(\ell_j) \ ,
\eeq
hence for instance
\bgroup
\def\x{0.6}
\tikzset{mynode/.append style={scale=\x}}
\tikzset{every picture/.append style={scale=\x, every node/.style=mynode, edge from parent/.style={myedge}, level 1/.style={sibling distance=10mm}, level 2/.style={sibling distance=5mm}, level distance=5mm}}
\beq
g^{(d+1)}_{\tikz{\node{} child{node{} child{node{}}} child{node{} child{node{}}};}}(N) = \frac{1}{\lambda \sqrt{2} }  \sum_{\substack{i,j=1\\i\neq j}}^l \mathcal{C}_1(\ell_i) \mathcal{C}_1(\ell_j)\ , \qquad
g^{(d+1)}_{\tikz{\node{} child{node{} child{node{}} child{node{}}} child{node{} child{node{}} child{node{}}};}}(N) = \frac{1}{\lambda \sqrt{2}} \sum_{\substack{i,j=1\\i\neq j}}^l\mathcal{C}_2(\ell_i) \mathcal{C}_2(\ell_j)\ .
\eeq
\egroup
If the root of $\gamma$ has exactly two offsprings, one being the root of one copy of $\mathfrak{t}_q$, the other of $\mathfrak{t}_{q'}$, with $q,q'>0$ and $q\neq q'$, then
\beq
g_\gamma^{(d+1)}(N) = \frac{1}{\lambda } \sum_{\substack{i,j=1 \\ i \neq j}}^l \mathcal{C}_q(\ell_i) \mathcal{C}_{q'}(\ell_j) \ ,
\eeq
in particular
\bgroup
\def\x{0.6}
\tikzset{mynode/.append style={scale=\x}}
\tikzset{every picture/.append style={scale=\x, every node/.style=mynode, edge from parent/.style={myedge}, level 1/.style={sibling distance=10mm}, level 2/.style={sibling distance=5mm}, level distance=5mm}}
\beq
g^{(d+1)}_{\tikz{\node{} child{node{} child{node{}} child{node{}}} child{node{} child{node{}}};}}(N) = \frac{1}{\lambda}\sum_{\substack{i,j=1\\i\neq j}}^l \mathcal{C}_1(\ell_i) \mathcal{C}_2(\ell_j) \ .
\eeq
\egroup
Finally, if the root of $\gamma$ has degree $r+1$, one of its offspring being the root of a copy of $\mathfrak{t}_q$ with $q>0$, while the other $r$ ones being without descendant, one has 
\beq
g_\gamma^{(d+1)}(N) = \frac{1}{\sqrt{\lambda}} \mathcal{C}_r(l-1) \sum_{i=1}^l \mathcal{C}_q(\ell_i) \ ,
\eeq
for instance
\bgroup
\def\x{0.6}
\tikzset{mynode/.append style={scale=\x}}
\tikzset{every picture/.append style={scale=\x, every node/.style=mynode, edge from parent/.style={myedge}, level 1/.style={sibling distance=10mm}, level 2/.style={sibling distance=5mm}, level distance=5mm}}
\beq
g^{(d+1)}_{\tikz{\node{} child{node{} child{node{}}} child{node{}};}}(N) = \frac{1}{\sqrt{\lambda}} \mathcal{C}_1(l-1) \sum_{i=1}^l \mathcal{C}_1(\ell_i) \ , \qquad
g^{(d+1)}_{\tikz{\node{} child{node{} child{node{}} child{node{}}} child{node{}};}}(N) = \frac{1}{\sqrt{\lambda}} \mathcal{C}_1(l-1) \sum_{i=1}^l \mathcal{C}_2(\ell_i) \ .
\eeq
\egroup

The inspection of this series of examples unveil some properties that are true in general, as we shall next prove:
\begin{enumerate}
\item $g_\gamma^{({d+1})}(N)$ depends on $N$ only through its first $d(\gamma)$ generations, where $d(\gamma)$ is the depth of the tree $\gamma$: on the examples when $d(\gamma)=0$ the function is constant, when $d(\gamma)=1$ it depends on $l$, and when $d(\gamma)=2$ it depends on $l,\ell_1,\dots,\ell_l$. Moreover $g_\gamma^{({d+1})}(N)$ does not depend on $d$ as long as $d+1 \ge d(\gamma)$, we shall therefore use the lighter notation $g_\gamma(N)$.
\item for $\gamma\neq\bullet$ one has $\E_0^{(d(\gamma))}[g_\gamma(N)]=0$, which follows from the orthogonality of the eigenvectors with $g_{\bullet}(N)=1$; this can be shown explicitly in the examples above using the orthonormality of the Charlier polynomials with respect to the Poisson distribution of parameter $\lambda$. A stronger property is actually true: for the trees $\gamma$ with $d(\gamma)=2$ one finds $\E_0^{(2)}[g_\gamma(N)|l]=0$ since, conditional on $l$, the $\ell_i$'s are i.i.d. Poissonian variables of mean $\lambda$, and since the Charlier polynomials $\mathcal{C}_q$ with $q>0$ have a zero average under this law.
\end{enumerate}

In order to generalize and formalize these properties let us introduce some additional notations. We shall write $\chi = \cup_{d \ge 0} \chi_d$ the set of rooted unlabeled finite trees (of arbitrary finite depth), and denote $\chi_\infty$ the set of unlabeled rooted locally finite trees (a tree in $\chi_\infty$ can thus have an infinite depth, but each of its vertices has a finite number of offsprings). We will denote $\mathbb{P}_0$ the probability law on $\chi_\infty$ associated to the Galton-Watson branching process with Poissonian offspring probability with parameter $\lambda$, and $\E_0$ the corresponding expectation. This means that the degree $l$ of the root of a tree $N$ drawn with the law $\mathbb{P}_0$ is a random variable with this Poisson distribution, and that $l$ i.i.d. samples of $\mathbb{P}_0$ are rooted at its offsprings. We will denote $\mathcal{F}_d$ the truncation of such an $N$ to its first $d$ generations; in technical terms it is the $\sigma$-algebra generated by the random variables that describe the first $d$ generations of $N$. These objects form a so-called filtration of the probability space, $\mathcal{F}_0\subset\mathcal{F}_1\subset\dots$, the first element $\mathcal{F}_0$ being the trivial $\sigma$-algebra bringing no information on the tree $N$.

We shall define $g_\gamma(N)$ for $\gamma\in\chi$ and $N\in\chi_\infty$ by the following formula, which is a slight variation of (\ref{eq:eigenvectors-alternative}):
\begin{align}
g_\gamma(N)&=\sqrt{\prod_{\beta\in\chi}\gamma_\beta!}\left[\prod_{\beta\in\chi}u_\beta^{\gamma_\beta}\right]e^{-\sqrt\lambda u_\bullet}\prod_{T\in\chi_\infty}\left(1+\sum_{\beta\in\chi}\frac{u_\beta}{\sqrt\lambda}g_\beta(T)\right)^{N_T} \label{eq_def_ggN_generic}\\
&=\sqrt{\prod_{\beta\in\chi}\gamma_\beta!}\left[\prod_{\beta\in\chi}u_\beta^{\gamma_\beta}\right]e^{-\sqrt\lambda u_\bullet}
\prod_{i=1}^l \left(1+\sum_{\beta\in\chi}\frac{u_\beta}{\sqrt\lambda}g_\beta(T_i)\right) \ , \label{eq_def_ggN_generic_2}
\end{align}
where $N$ is defined as $\{N_T\}_{T\in\chi_\infty}$ in the first line or as a list of subtrees $(T_1,\dots,T_l)$ arbitrarily ordered in the second line. We claim that this formula is well-defined and fixes unambiguously the value of $g_\gamma(N)$ by induction on $d(\gamma)$. As a matter of fact if $d(\gamma)=0$ then $\gamma=\bullet$, the trivial tree with $\gamma_\beta=0$ for all $\beta\in\chi$, and then (\ref{eq_def_ggN_generic}) implies $g_\bullet(N)=1$. For the induction step assume that the $g_\beta$ have been defined for all $\beta$ with $d(\beta)\le d$; consider then a tree $\gamma$ with $d(\gamma)=d+1$. The $\beta$'s such that $\gamma_\beta >0$ have $d(\beta) \le d$, hence the right-hand side of (\ref{eq_def_ggN_generic}) only involves quantities that have been already defined along the induction.

It will be convenient in the following to distinguish in a dual tree $\gamma$ the offsprings of the root which have a non-trivial progeny from the others. We shall thus denote, for $\gamma\in\chi$, $\gamma_{\rm nt}=\sum_{\beta\neq\bullet}\gamma_\beta$ the number of non-trivial branches emerging from its root. Let us also distinguish the trivial trees in (\ref{eq_def_ggN_generic_2}) and write:
\begin{align}
    \prod_{i=1}^l\left(1+\sum_{\beta\in\chi}\frac{u_\beta}{\sqrt\lambda}g_\beta(T_i)\right)
    & =\prod_{i=1}^l\left(1+\frac{u_\bullet}{\sqrt\lambda}+\sum_{\beta\neq\bullet}\frac{u_\beta}{\sqrt\lambda}g_\beta(T_i)\right)\\
    & =\sum_{I\subset[l]}\left(1+\frac{u_\bullet}{\sqrt\lambda}\right)^{l-|I|}\prod_{i\in I}\left(\sum_{\beta\neq\bullet}\frac{u_\beta}{\sqrt\lambda}g_\beta(T_i)\right)\\
    &=\sum_{I\subset[l]}\left(1+\frac{u_\bullet}{\sqrt\lambda}\right)^{l-|I|}\frac{1}{\lambda^{|I|/2}}\sum_{\{\beta_i\neq\bullet\}_{i\in I}}\prod_{i\in I}u_{\beta_i}g_{\beta_i}(T_i) \ .
\end{align}
It thus follows from (\ref{eq_def_ggN_generic_2}) that
\begin{equation}
    g_\gamma(N) = \sum_{I\subset[l]} \left\{ \left(\sqrt{\gamma_\bullet!}[u_\bullet^{\gamma_\bullet}]e^{-\sqrt\lambda u_\bullet}\left(1+\frac{u_\bullet}{\sqrt\lambda}\right)^{l-|I|}\right)
    \frac{1}{\lambda^{|I|/2}}\sum_{\{\beta_i\neq\bullet\}_{i\in I}}\prod_{i\in I}g_{\beta_i}(T_i)
    \sqrt{\prod_{\beta\neq\bullet}\gamma_\beta!}\left[\prod_{\beta\neq\bullet}u_\beta^{\gamma_\beta}\right]\prod_{i\in I}u_{\beta_i}
    \right\} \ . \label{eq_def_ggN_generic_decomposed}
\end{equation}
We underline that the term $\left[\prod_{\beta\neq\bullet}u_\beta^{\gamma_\beta}\right]\prod_{i\in I}u_{\beta_i}$ is non-zero if and only if the $\{\beta_i\neq\bullet\}_{i\in I}$ are a permutation of the $\{\beta:\gamma_\beta>0,\beta\neq\bullet\}$, counted with their multiplicities $\gamma_\beta$, which implies that $|I|=\gamma_{\rm nt}$.

From this expression we can prove the following properties by induction on $d(\gamma)$:
\begin{itemize}
    \item[1.] $g_\gamma(N)$ depends only on the $d(\gamma)$ first generations of $N$, or in technical terms $g_\gamma$ is $\mathcal{F}_{d(\gamma)}$-measurable. As a matter of fact,
    \begin{itemize}
        \item if $d(\gamma)=1$ then $\beta_{\rm nt}=0$, hence
        \begin{equation}
        g_\gamma(N)=\sqrt{\gamma_\bullet!}[u_\bullet^{\gamma_\bullet}]e^{-\sqrt\lambda u_\bullet}\left(1+\frac{u_\bullet}{\sqrt\lambda}\right)^l = \mathcal{C}_{\gamma_\bullet}(l) \ ,
        \end{equation}
the Charlier polynomial defined in Eq.~(\ref{eq_def_Charlier}). This depends on $N$ only through the degree $l$ of its root, in other words it is $\mathcal{F}_1$-measurable;
\item assume that $g_\beta$ is $\mathcal{F}_{d(\beta)}$-measurable for all $\beta\in \chi_d$, and consider $\gamma$ with $d(\gamma)=d+1$. The non-zero elements of $\{\gamma_\beta\}$ correspond to trees $\beta$ with $d(\beta) \le d$, hence in the expression (\ref{eq_def_ggN_generic_decomposed}) of $g_\gamma$ the functions $g_{\beta_i}(T_i)$ that appear in the right-hand side depend on $T_i$ only through their first $d(\beta_i) \le d$ generations by the induction assumption. As a consequence $g_\gamma(N)$ only depends on the $d+1=d(\gamma)$ first generations of $N$, concluding the induction step.
    \end{itemize}
    \item[2.]
    $\E_0[g_\gamma(N)|\mathcal{F}_d]=\begin{cases}g_\gamma(N)&\text{if $d\geq d(\gamma)$}\\0&\text{if $d<d(\gamma)$}\end{cases}$ .\\
    The first case is actually a direct consequence of the property 1. we just proved. We shall justify the vanishing of the conditional expectation in the second case by induction on $d(\gamma)$:
    \begin{itemize}
        \item if $d(\gamma)=1$, then $d<d(\gamma)$ means $d=0$, hence $\E_0[g_\gamma(N)|\mathcal{F}_0]=\E_0[g_\gamma(N)]$. Furthermore, the trees $\gamma$ of depth $d(\gamma)=1$ are of the form $\gamma=\mathfrak{t}_q$ for some $q>0$. As a consequence $g_\gamma(N)=\mathcal{C}_q(l)$ is a Charlier polynomial orthogonal to $\mathcal{C}_0(l)=1$ under the Poisson law of parameter $\lambda$, which implies $\E_0[g_\gamma(N)|\mathcal{F}_0]=0$.
        \item consider now a tree $\gamma$ with $d(\gamma)>1$, a depth $d<d(\gamma)$, and assume that the claim has been proven for all previous induction steps. Since $d(\gamma)>1$ one has $\gamma_{\rm nt} \neq 0$, hence the terms that contribute to (\ref{eq_def_ggN_generic_decomposed}) contain a non-empty product $\prod_{i\in I}g_{\beta_i}(T_i)$, where the $\beta_i$ are in correspondence with $\{\beta:\gamma_\beta>0,\beta\neq\bullet\}$. Since $d(\gamma)=1+\max \{d(\beta) : \gamma_\beta>0\}$, there is at least one index $i_0 \in I$ such that $d(\beta_{i_0})=d(\gamma)-1$. By the decomposition of $N$ as a root and $l$ subtrees $T_1,\dots,T_l$, conditioning $N$ on its first $d$ generations implies to condition $T_{i_0}$ on its first $d-1$ generations. Since $d-1<d(\beta_{i_0})$ and since $T_{i_0}$ is also drawn from $\mathbb{P}_0$ one can use the induction assumption to obtain $\E_0[ g_{\beta_{i_0}}(T_{i_0}) |\mathcal{F}_d]=0$. Noting finally that conditional on $l$ the subtrees $T_i$ are independent concludes the justification of the induction step, $\E_0[g_\gamma(N)|\mathcal{F}_d]=0$ for $d<d(\gamma)$.
\end{itemize}
\end{itemize}

These results can be summarized as
\beq
\E_0[g_\gamma(N)|\mathcal{F}_d] = g_\gamma(N) \ \one(d(\gamma) \le d) \ ,
\label{eq_identity_conditional}
\eeq
an identity that will finally be used to unveil a property of the approximated likelihood ratios $\hL^{(d)}$. Consider indeed a pair $(N,N')$ of independent Galton-Watson trees, i.e. drawn from $\mathbb{P}_0 \otimes \mathbb{P}_0$ (with averages still denoted $\E_0$), and call $\mathcal{F}_d$ the $\sigma$-algebra corresponding to the information contained in the first $d$ generations of $N$ and $N'$. Let $\hchi \subset\chi$ be an arbitrary family of trees and define
\begin{equation}
\hL^{(d)}(N,N')=\sum_{\gamma\in\hchi}w_\gamma \, g_\gamma(N) g_\gamma(N')\ \one(d(\gamma)\leq d) \ ,
\end{equation}
for some arbitrary coefficients $w_\gamma$ (sufficiently small for the sums to converge). This notation coincides with the one used in the rest of the text if one takes for $\hchi$ the trees in which each vertex has at most $m$ offsprings, and $w_\gamma= s^{|\gamma|-1}$. Thanks to the properties of the $g_\gamma$ established previously, $\hL^{(d)}$ is $\mathcal{F}_d$-measurable. Moreover,
\begin{equation}
    \E_0[\hL^{(d+1)} (N,N')|\mathcal{F}_d]=\sum_{\gamma\in\hchi} w_\gamma 
\E_0[g_\gamma(N)|\mathcal{F}_d] \E_0[g_\gamma(N')|\mathcal{F}_d]
\ \one(d(\gamma)\leq d+1) = \hL^{(d)}(N,N') \ ,
\end{equation}
using (\ref{eq_identity_conditional}) to simplify the conditional expectation. This shows that $\hL^{(d)}$, viewed as a sequence of random variables indexed by $d$, is a martingale under $\mathbb{P}_0 \otimes \mathbb{P}_0$. This martingale property was established for the likelihood ratio $L^{(d)}$ in~\cite{ganassali_correlation_2022}, and played an important role in the proof of the connection between the alignment of graphs and the hypothesis testing problem on trees. One can hope that the martingale character of $\hL^{(d)}$ may help to adapt these reasonings to the finite $m$ message-passing algorithms (even if $\hL^{(d)}$ is not positive anymore, hence some martingale convergence theorems cannot be applied directly).

\subsection{Simplifications in the large degree limit}
\label{app_largedeg}

We explain in this Appendix the simplifications that arise in the $\lambda \to \infty$ limit and that were mentioned in Sec.~\ref{sec:trees-results}. It is instructive to start with an elementary computation about the hypothesis testing problem where one observer is provided with a pair of random variables $(z,z') \in \mathbb{R}^2$ and has to decide whether they are drawn from a law $\mathbb{P}_0$ in which they are independent or from a law $\mathbb{P}_1$ under which they are correlated. More precisely, both laws are assumed to be Gaussian and thus characterized by their first two moments:
\beq
\E_0[z]=\E_0[z']=\E_1[z]=\E_1[z']=0 \ , \ \ \E_0[z^2]=\E_0[z'^2]=\E_1[z^2]=\E_1[z'^2]=1 \ , \ \ \E_0[z z'] = 0 \ , \ \ \E_1[z z'] = w \ ,
\eeq
where $w \in [0,1] $ is the correlation coefficient under $\mathbb{P}_1$. From the knowledge of these moments the corresponding probability densities $p_0(z,z')$ and $p_1(z,z')$ are easily written. Let us denote $L_w^{\rm G}(z,z')=\frac{p_1(z,z')}{p_0(z,z')}$ the likelihood ratio; one can obtain its diagonalization with the following manipulations:
\begin{align}
  L_w^{\rm G}(z,z') & = \frac{1}{\sqrt{1-w^2}} \exp\left( - \frac{1}{2} \begin{pmatrix} z & z' \end{pmatrix} \begin{pmatrix} 1 & w \\ w & 1 \end{pmatrix}^{-1} \begin{pmatrix} z \\ z' \end{pmatrix} + \frac{1}{2} z^2 + \frac{1}{2} z'^2  \right) \label{eq_def_LG} \\
  & = \int \frac{{\rm d} x {\rm d} x'}{2 \pi} \ \exp\left(  - \frac{1}{2} \begin{pmatrix} x & x' \end{pmatrix} \begin{pmatrix} 1 & w \\ w & 1 \end{pmatrix} \begin{pmatrix} x \\ x' \end{pmatrix} + i \begin{pmatrix} x & x' \end{pmatrix} \begin{pmatrix} z \\ z' \end{pmatrix} + \frac{1}{2} z^2 + \frac{1}{2} z'^2 \right) \\
          & = \int \frac{{\rm d} x {\rm d} x'}{2 \pi} \ e^{w (-ix)(-ix')}
            \ e^{-\frac{1}{2} x^2 + ixz + \frac{1}{2} z^2 }
            \ e^{-\frac{1}{2} x'^2 + ix'z' + \frac{1}{2} z'^2 } \\
          & = \sum_{q=0}^\infty w^q \ {\cal H}_q(z) {\cal H}_q(z') \ , \label{eq_LG_expansion}
\end{align}
where in the second line we performed a so-called Hubbard-Stratonovich transformation to decouple the dependency on $z$ and $z'$, the price to pay being the introduction of additional Gaussian integrals on $x$ and $x'$. In the last line we have defined
\begin{align}
  {\cal H}_q(z) & = \frac{1}{\sqrt{q!}} \int \frac{{\rm d}x}{\sqrt{2 \pi}} (-ix)^q  e^{-\frac{1}{2} x^2 + ixz + \frac{1}{2} z^2 } \\
                & = \sqrt{q!} [u^q] \int \frac{{\rm d}x}{\sqrt{2 \pi}}  e^{-\frac{1}{2} x^2 + ix(z-u) + \frac{1}{2} z^2 } \\
  & = \sqrt{q!} [u^q] e^{-\frac{1}{2} u^2 + u z} \ , \label{eq_def_Hermite}
\end{align}
where one recognizes through the last form that ${\cal H}_q$ is the $q$-th Hermite polynomial. Those form the sequence of orthonormal polynomials for the standard Gaussian measure, and read for the smallest values of $q$:
\beq
{\cal H}_0(z) = 1 \ , \qquad 
{\cal H}_1(z) = z \ , \qquad 
{\cal H}_2(z) = \frac{z^2-1}{\sqrt{2}} \ .
\eeq
With elementary computations starting from the definition (\ref{eq_def_LG}) one can obtain
\beq
\E_1[ L_w^{\rm G} ] = \frac{1}{1-w^2} \ , \qquad \E_1[ \ln L_w^{\rm G} ] = -\frac{1}{2} \ln(1-w^2) \ ,
\eeq
hence the relationship $\E_1[ \ln L_w^{\rm G} ] = \frac{1}{2} \ln \E_1[ L_w^{\rm G} ]$ for this simple finite-dimensional problem.

Let us turn now to the problem of trees, and the simplifications it undergoes in the large degree limit. Recall that a tree $N \in \chi_{d+1}$ is described by the vector $N=\{N_T\}_{T \in \chi_d}$ of non-negative integers counting the number of copies of subtrees rooted at the offsprings of the root of $N$. One can equivalently introduce a real vector $y=\{y_\beta\}_{\beta \in \chi_d}$ which is a centered, normalized and rotated version of $N$, namely
\beq
y_\beta= \frac{1}{\sqrt{\lambda}} \sum_T g_\beta^{(d)}(T) (N_T - \lambda \pzero{d}{T}) \ .
\eeq
The orthogonality properties of the eigenvectors $g$ ensure the invertibility of this transformation, the coordinates of $N$ can be expressed in terms of those of $y$ as (see~\cite{ganassali_statistical_2022} for more details):
\beq
N_T = \lambda \pzero{d}{T} + \sqrt{\lambda} \pzero{d}{T} \sum_\beta y_\beta g_\beta^{(d)}(T) \ .
\label{eq_NT_from_y}
\eeq
The advantage of this representation in terms of $y$ rather than in terms of $N$ appears in the large $\lambda$ limit: the coordinates $y_\beta$ remain indeed finite in this limit. Inserting (\ref{eq_NT_from_y}) in the expression (\ref{eq:eigenvectors-alternative}) of the recursion on the eigenvectors and expanding in the limit $\lambda \to \infty$, one obtains after a short computation
\begin{align}
  g_\gamma^{(d+1)}(y)& = \sqrt{\prod_\beta \gamma_\beta ! } [u^\gamma] \exp \left( - \sqrt{\lambda} \sum_{\beta,T} u_\beta \pzero{d}{T} g_\beta^{(d)}(T) \right. \\ &
 \hspace{3cm} \left. + \sum_T \left( \lambda \pzero{d}{T} + \sqrt{\lambda} \pzero{d}{T} \sum_\beta y_\beta g_\beta^{(d)}(T)\right) \ln \left(1 + \frac{1}{\sqrt{\lambda}} \sum_\beta u_\beta g_\beta^{(d)}(T)  \right) \right)
\nonumber \\
& = \sqrt{\prod_\beta \gamma_\beta ! } [u^\gamma] \exp\left(-\frac{1}{2} \sum_\beta u_\beta^2 + \sum_\beta u_\beta y_\beta + O\left(\frac{1}{\sqrt{\lambda}} \right) \right) \\
& = \prod_{\beta \in \chi_d} {\cal H}_{\gamma_\beta}(y_\beta) \left(1 + O\left(\frac{1}{\sqrt{\lambda}} \right)\right) \ , \label{eq_g_largel}
\end{align}
where we recognized in the last line the definition of the Hermite polynomials given in (\ref{eq_def_Hermite}). We can thus obtain the large degree limit of the likelihood ratio between two trees of depth $d+1$ by inserting this expression in the diagonalization formula (\ref{eq:theL2formula}); keeping implicit the limit $\lambda \to \infty$, this gives:
\begin{align}
  L^{(d+1)}(y,y') & = \sum_{\gamma \in \chi_{d+1}} s^{|\gamma|-1}  g_\gamma^{(d+1)}(y)g_\gamma^{(d+1)}(y') \\
                  & = \prod_{\beta \in \chi_d} \sum_{\gamma_\beta =0}^\infty (s^{|\beta|})^{\gamma_\beta} {\cal H}_{\gamma_\beta}(y_\beta) {\cal H}_{\gamma_\beta}(y'_\beta) \\
  & = \prod_{\beta \in \chi_d} L_{s^{|\beta|}}^{\rm G}(y_\beta,y'_\beta) \ ,
\label{eq_L_largedeg}
\end{align}
using the expansion of the likelihood ratio for pairs of Gaussian established in (\ref{eq_LG_expansion}). The infinite dimensional likelihood ratio for pairs of trees thus factorizes into a product of finite-dimensional objects; as a matter of fact it was shown in~\cite{ganassali_statistical_2022} that the vector $(y,y')$ converges in law, in the large $\lambda$ limit, to a Gaussian vector under both $\pone{d+1}{}$ and $\pzero{d+1}{} \otimes \pzero{d+1}{}$, with such a block structure for its covariance matrix. This justifies the statement (\ref{eq_E1L1_Gaussian}) made in the main text, as a direct consequence of the factorization and of the property $\E_1[ \ln L_w^{\rm G} ] = \frac{1}{2} \ln \E_1[ L_w^{\rm G} ]$ established above by direct computations in the finite-dimensional Gaussian case.

Consider now the finite $m$ situation and the large $\lambda$ behavior of $\hL^{(d+1)}$; unfortunately we do not obtain a description as simple as in the $m=\infty$ case. Inserting the large $\lambda$ formula (\ref{eq_g_largel}) of $g_\gamma^{(d+1)}$ in the definition (\ref{eq_def_hLd}) one obtains
\beq
\hL^{(d+1)}(y,y') = \sum_{\gamma \in \hchi_{d+1}} s^{|\gamma|-1} \prod_{\beta \in \hchi_d}  {\cal H}_{\gamma_\beta}(y_\beta) {\cal H}_{\gamma_\beta}(y'_\beta) \ ,
\eeq
but the degree constraints in the definition of $\hchi_{d+1}$ do not allow to fully perform the resummation as in (\ref{eq_L_largedeg}). A compact expression can be obtained by using a variant of identity (\ref{eq_summation_tchi}) to perform the summation over $\hchi_{d+1}$, yielding
\beq
\hL^{(d+1)}(y,y') = \sum_{p=0}^m [t^p] \prod_{\beta \in \hchi_d} L_{t s^{|\beta|}}^{\rm G}(y_\beta,y'_\beta) \ ,
\eeq
but it remains quite a challenge to extract from this expression the typical behavior of $\hL^{(d+1)}$ under the law $\pone{d+1}{}$.

\section{Otter's constants}\label{app:modified-otter}

We give in this Appendix some additional details on the enumeration of various families of trees, justifying part of the results that were used in Sec.~\ref{sec:trees-results}. We recall that
\begin{equation}
A_{d,n}=|\{T\in\chi_{d},|T|=n\}|
\end{equation}
denotes the number of unlabeled rooted trees of depth at most $d$ containing $n$ vertices. For a fixed value of $n$ this quantity grows in a monotonic way with $d$, since increasing $d$ corresponds to releasing a constraint ($\chi_d \subset \chi_{d+1}$). We denote $A_n$ its limit as $d\to \infty$, which is actually reached for $d \ge n-1$, since a tree of $n$ vertices has depth at most $n-1$. We introduce the associated generating functions, 
\beq
\psi_d(x) =\sum_{n\geq1} A_{d,n} x^{n-1} \ , \qquad \psi(x) =\sum_{n\geq1} A_n x^{n-1} \ ,
\eeq
which contain the same information than the sequences indexed by $n$, but can be manipulated in a more convenient way. Let us derive indeed an induction relation between the $\psi_d$'s, recalling that a tree $N \in \chi_{d+1}$ is specified by the numbers $\{N_T\}_{T \in \chi_d}$ of the copies of trees of depth at most $d$ rooted at the offsprings of its root, and that in this decomposition $|N| = 1 + \sum_T |T| N_T$. This yields
\begin{align}
\psi_{d+1}(x) & = \sum_{N \in \chi_{d+1}} x^{|N|-1} 
= \sum_{\{N_T \ge 0\}_{T \in \chi_d}} \prod_{T \in \chi_d} x^{|T|N_T} 
= \prod_{T \in \chi_d} \frac{1}{1-x^{|T|}}  \\
& = \exp\left( - \sum_{T\in \chi_d} \log(1-x^{|T|} ) \right) 
 = \exp\left( \sum_{j=1}^\infty \frac{1}{j} \sum_{T\in \chi_d} x^{j|T|} \right) \\
& = \exp\left( \sum_{j=1}^\infty \frac{1}{j} x^j \psi_d(x^j) \right) \ .
\end{align}
This functional recursion between $\psi_d$ and $\psi_{d+1}$ implies, taking the limit $d\to\infty$, the following functional fixed-point equation on the generating function $\psi(x)$ of the $A_n$:
\beq
\psi(x) = \exp\left( \sum_{j=1}^\infty \frac{1}{j} x^j \psi(x^j) \right) \ .
\eeq
With more work~\cite{otter_number_1948} one can show that the radius of convergence of this power series is strictly positive and finite; it is denoted $\alpha$ and called Otter's constant. A numerical study~\cite{otter_number_1948} yields $\alpha \approx 0.338$. The value of the radius of convergence of a power series is linked to the large $n$ behavior of its coefficients, at the exponential scale. One can furthermore show that $\psi(\alpha)$ is finite, with a square root behavior as $x \to \alpha^-$. This implies the following asymptotic behavior for the $A_n$:
\beq
A_n \underset{n\to \infty} \sim \frac{C}{n^{3/2}} \left(\frac{1}{\alpha}\right)^n \ ,
\eeq
for some constant $C$.

Consider now the corresponding counting problem in the constrained set of trees $\hchi_d \subset \chi_d$ where all vertices have a number of offsprings smaller or equal to $m$. We denote $\hA_{d,n}$, $\hA_n$, $\hpsi_d$ and $\hpsi$ the quantities corresponding to $A_{d,n}$, $A_n$, $\psi_d$ and $\psi$ with this additional constraint (they depend of course on $m$, even if the notation does not make this explicit). One can amend the recursion on $\psi_d$, by imposing that the non-zero values of $N_T$ correspond to $T \in \hchi_d$, and enforcing the constraint $\sum_T N_T \le m$. This is precisely the formula used in (\ref{eq_summation_tchi}), with $s$ replaced by $x$, $\gamma$ by $N$, and $h=1$. This gives:
\begin{align}
\hpsi_{d+1}(x) & = \sum_{p=0}^m [t^p] \sum_{\{N_T \ge 0\}_{T \in \hchi_d}} \prod_{T \in \hchi_d} (t x^{|T|})^{N_T} 
= \sum_{p=0}^m [t^p] \prod_{T \in \hchi_d} \frac{1}{1-t x^{|T|}}  \nonumber \\
& = \sum_{p=0}^m [t^p] \exp\left( - \sum_{T\in \hchi_d} \log(1- t x^{|T|} ) \right) 
 = \sum_{p=0}^m [t^p] \exp\left( \sum_{j=1}^\infty \frac{1}{j} t^j \sum_{T\in \hchi_d} x^{j|T|} \right) \nonumber \\
& = \sum_{p=0}^m [t^p] \exp\left( \sum_{j=1}^\infty \frac{1}{j} t^j x^j \hpsi_d(x^j) \right) \ . \label{eq_psidp1}
\end{align}

For $m=2$, one can be more explicit and the last equation becomes:
\beq
\hpsi_{d+1}(x) = 1 + x \hpsi_d(x) + \frac{1}{2} x^2 \hpsi_d(x^2) + \frac{1}{2} x^2 \hpsi_d(x)^2 \ . \label{eq_psidp1_m2}
\eeq
This has a direct combinatorial interpretation: a tree in $\hchi_{d+1}$ is either empty, or its root has degree 1 and its offspring is the root of a tree in $\hchi_d$, or its root has degree 2, with two subtrees $T,T' \in \hchi_d$; in the latter case one has to be careful in the counting of distinct trees, and in particular take into account whether $T=T'$ or not. For $m=3$ one finds similarly
\beq
\hpsi_{d+1}(x) = 1 + x \hpsi_d(x) + \frac{1}{2} x^2 \hpsi_d(x^2) + \frac{1}{2} x^2 \hpsi_d(x)^2  + \frac{1}{3} x^3 \hpsi_d(x^3)
+ \frac{1}{2} x^3 \hpsi_d(x) \hpsi_d(x^2) + \frac{1}{6} x^3 \hpsi_d(x)^3 \ . \label{eq_psidp1_m3}
\eeq
Taking the limit $d \to \infty$ (in (\ref{eq_psidp1}) for the generic $m$ case, or in (\ref{eq_psidp1_m2},\ref{eq_psidp1_m3}) for $m=2$ and $m=3$ respectively) yields a functional fixed-point equation on $\hpsi$, the generating function of the $\hA_n$. The $m$-dependent Otter's constant $\halpha$ can then be defined as the radius of convergence of $\hpsi$, or equivalently as the inverse of the exponential growth rate of the $\hA_n$. These constants were  evaluated numerically in~\cite{otter_number_1948} for $m=2$ and $m=3$, yielding respectively $\halpha \approx 0.403$ and $\halpha \approx 0.355$. The $\halpha$'s decrease monotonically towards $\alpha$ when $m$ increases, since this modification of the parameter $m$ corresponds to the inclusion of more and more trees in the sets $\hchi_d$. The analytic determination of the asymptotic behavior of $\halpha$ when $m$ diverges seems challenging, since they are defined as the radius of convergence of functions solution of an implicit functional equation. From the numerical determination of $\halpha$ with $m\in\{2,\dots,6\}$ we found a rather good fit of the form $\halpha \approx \alpha + a \, e^{- b \, m}$ with $a$ and $b$ two fitting parameters, this exponential convergence with $m$ of the modified Otter's constants towards $\alpha$ would thus be a reasonable conjecture for their asymptotic behavior as $m\to\infty$.

\end{document}